\documentclass[aps,preprint,showpacs,floatfix]{revtex4}
\pdfoutput=1
\usepackage{graphicx,bm,amsmath,dcolumn}
\RequirePackage{color}

\definecolor{MyDarkGreen}{rgb}{0.02,0.60,0.06}

\begin{document}
	\title{Exact finite-size corrections in the dimer model on a cylinder.}
	\author{Vladimir V. Papoyan}
	\email{vpap@theor.jinr.ru}
		\affiliation{Bogoliubov Laboratory of Theoretical Physics, Joint Institute for Nuclear Research, 141980 Dubna, Russian Federation}
	\affiliation{Dubna State University, 141982 Dubna, Russian Federation}
	
	\begin{abstract}	
The exact finite-size corrections to the free energy $F$ of the dimer model on lattice $\mathcal{M} \times \mathcal{N}$ with cylindrical boundary
conditions have been derived for three cases where the lattice is completely covered by dimers: $\mathcal{M} = 2M$, $\mathcal{N} = 2N$; $\mathcal{M} = 2M - 1$, $\mathcal{N} = 2N$; and $\mathcal{M} = 2M$, $\mathcal{N} = 2N - 1$. For these types of cylinders, ratios $r_p(\rho)$ of the $p$th coefficient of $F$ have been calculated for the infinitely long cylinder (${\mathcal M} \rightarrow \infty$) and infinitely long strip (${\mathcal N} \rightarrow \infty$) at varying aspect ratios.
As in previous studies of the dimer model on the rectangular lattice with free boundary conditions and for the Ising model with Brascamp-Kunz  boundary conditions, the limiting values $p \to \infty$ exhibit abrupt anomalous behaviour of ratios $r_p(\rho)$ at certain values of $\rho$. These critical values of $\rho$ and the limiting values of the finite-size expansion coefficient ratios vary between the different models.
	\end{abstract}		
		
	\maketitle

	\pagebreak

\section{Introduction}

In systems with a finite size, it is essential to consider the boundaries, regardless of their significance in the thermodynamic limit. Finite-size scaling analysis allows us to identify critical and non-critical features in finite systems and extends our analysis to infinite systems.  This approach was presented in the groundbreaking work of Fisher and Barber in 1972 \cite{FisherBarber}, which served as the basis for analyzing thermodynamic properties on all scales.

In accordance with the concept of finite-size scaling, various models were analyzed using both analytical and Monte Carlo simulation methods on different lattice structures with varying boundary conditions for finite systems  \cite{Privman1990}. In the context of this study, the following models can be mentioned: Ising models  \cite{Salas, Kenna, Izmailian2002, Izmailian2002a, Izmailian2007a, Izmailian2012a}, dimer models \cite{Kasteleyn, FisherTemp, Brankov, Izmailian2002, Izmailian2003, Izmailian2006, Izmailian2014, Izmailian2016, Nigro2012, Allegra2014, Allegra2015, Kenyon2016, Chhita2016, Pearce2017, Bleher2018, Izmailian2019}, Gaussian models \cite{Izmailian2002}, spanning tree models \cite{Izmailian2015a}, resistor network models \cite{Izmailian2010, Izmailian2014a}, and percolation models \cite{Hu1, Hu2, Hu3, Ziff, ZiffKleb,ZiffLor,ZifFur, Mal}.

The Ising and dimer models in low dimensions are notable examples of rare cases where analytical or exact solutions can be found. These analytical solutions are significant for determining precise scaling forms for finite systems, understanding the impact of boundary conditions, and eliminating common errors in simulations.

Kasteleyn \cite{Kasteleyn}, as well as Temperley and Fisher \cite{FisherTemp} in the 1960s, found the first few coefficients of the asymptotic expansion of the free energy of the dimer model on a rectangular $2M \times 2N$ lattice with free boundary conditions. This remains one of the most significant exact results in fundamental statistical mechanics. A recent study by Izmailian, Papoyan, and Ziff \cite{Izmailian2019} on the finite-size  corrections in the dimer model, was inspired by these works and followed the approach outlined in \cite{Izmailian2002} and \cite{Izmailian2003}. They obtained exact expressions for these coefficients up to the 22nd order in terms of elliptic theta functions $\theta_2, \theta_3, \theta_4$ and the elliptic integral of the second kind $E$.

One of the main findings of this study was an intriguing result: the asymptotic coefficients for infinitely long strip and square, for large orders, approached exactly a factor of $1/2$ in ratio. In contrast, the ratio between asymptotic coefficients of infinite strips and rectangles approached $1$ as the order increased, when scaled by aspect ratio. This observation provided a better understanding of coefficients in the asymptotic behavior of these shapes, as the simpler form of strip coefficients made them easier to analyze.

To further investigate this unexpected behavior, the authors recommended examining other available models. The Ising model with Brascamp-Kunz boundary conditions, studied by Izmailian, Kenna and Papoyan  \cite{Izmailian2023}, emerged as one such alternative. Their research yielded  exact finite-size corrections for the free energy $F$ of the Ising model on the ${\cal M} \times 2 {\cal N}$ square lattice with Brascamp-Kunz boundary conditions. The ratios $r_p(\rho)$ of $p$th coefficients of $F$ for the square lattice to coefficients of $F$ for the infinitely long cylinder (${\cal M} \to \infty$) and to coefficients of $F$ for the infinitely long Brascamp-Kunz strip (${\cal N} \to \infty$) at varying values of the aspect ratio  $\rho={(\cal M}+1) / 2{\cal N}$  were calculated as well. Similarly to previous studies for the dimer model on a rectangular $2M \times 2N$ lattice with free boundary conditions, the limiting values of $p \rightarrow \infty$ for $r_p(\rho)$ exhibit abrupt anomalous behavior at certain values of $\rho$. For increasing $p$, the values of the ratios $r_p(\rho)$ tend to $1$ for $\rho\neq 1/2$, while they vanish exactly at $\rho = 1/2$ in both infinitely long cylinder and infinitely long Brascamp-Kunz strip cases. However, critical values of $\rho$ and the limiting values of the  finite-size-expansion coefficient ratios differ, between dimer model on the rectangular and Ising model with Brascamp–Kunz boundary conditions.

The aim of the present paper is to gain a better understanding of the abrupt change in the behavior of the $r_p(\rho)$ ratio by examining it in the dimer model with cylindrical boundary conditions.

\section{Dimer model on cylinder lattice}

The partition function of the dimer model on $\mathcal{M}\times\mathcal{N}$ lattice is given by
\begin{equation}
Z_{\mathcal{M},{\mathcal{N}}}(z_v, z_h)=\sum z_v^{n_v} z_h^{n_h}
\label{Zdimer}
\end{equation}
where the summation is taken over all possible dimer covering configurations, $z_h$ and $z_v$ are the dimer weight in the horizontal and vertical directions respectively, and $n_v$ and $n_h$ are the number of vertical and horizontal dimers respectively  \cite{Kasteleyn}.

The explicit expression of the partition
function depends crucially on whether $\mathcal{M}$ and $\mathcal{N}$ are even or
odd, and since the total number of sites must be even if the
lattice is to be completely covered by dimers, we will consider
three cases $\mathcal{M} = 2M$, $\mathcal{N} = 2N$; $\mathcal{M} = 2M - 1$, $\mathcal{N} = 2N$; $\mathcal{M} = 2M$, $\mathcal{N} = 2N - 1$.

In accordance with results for isotropic case ($z_v=z_h=z=1$) from  \cite{Izmailian2003}, Eq.\  (\ref{Zdimer}) can be written in terms of the partition function with twisted
boundary conditions $Z_{\alpha,\beta}(M,N)$  which is defined as follows:
\begin{eqnarray}
Z_{\alpha,\beta}^{2}(M,N) &=& \prod_{n = 0}^{N - 1} \prod_{m = 0}^{M - 1}4\left[\sin^{2}\left(\frac{\pi(n + \alpha)}{N}\right) + \sin^{2}\left(\frac{\pi(m + \beta)}{M}\right)\right].
\label{twistz}
\end{eqnarray}
Then, for the three cases of dimers under consideration, the results of the partition function on cylinder can be reduced to:
\begin{eqnarray}
Z_{2M,2N}^\mathrm{cyl}(1)&=&\frac{
	Z_{\frac{1}{2},\frac{1}{2}}(2M + 1, N)}
{2 \cosh{ \left(N {\rm arcsinh} 1\right)}}	
\label{cyl1}
\end{eqnarray}
for dimers on $2M \times 2N$ cylinder,
\begin{eqnarray}
Z_{2M - 1,2N}^\mathrm{cyl}(1)&=&\frac{
	Z_{\frac{1}{2},0}(2M, N)}
{2 \cosh{ \left(N {\rm arcsinh} 1\right)}}	
\label{cyl2}
\end{eqnarray}
for dimers on $(2M - 1) \times 2N$ cylinder,
\begin{eqnarray}
Z_{2M,2N - 1}^\mathrm{cyl}(1)&=&\left[\frac{
	Z_{0,\frac{1}{2}}(2M + 1, 2N - 1)}
{2 \cosh{ \left((2N - 1) {\rm arcsinh} 1\right)}}\right]^\frac{1}{2}	
\label{cyl3}
\end{eqnarray}
for dimers on $2M \times (2N - 1)$ cylinder.

\section{Finite-size scaling theory}

The research by Ivashkevich, Izmailian, and Hu \cite{Izmailian2002} presents a systematic approach to calculating finite-size corrections to the partition function for free models on a torus, particularly focusing on the dimer model on $2M\times2N$ lattice. They also examined the Ising and Gaussian models on $M\times N$ lattices. The authors obtained all terms in the exact asymptotic expansion of the logarithm of the partition function for a class of free, exactly solvable statistical mechanics models.
This method is rooted in the deep connection between asymptotic expansion terms and Kronecker's double series. In later studies, Izmailian, Oganesyan, and Hu \cite{Izmailian2003} expanded this framework to the dimer model on rectangular $M\times N$ lattices, exploring various boundary conditions and the impact of lattice site parity. In subsequent work by Izmailian et al. \cite{Izmailian2014} were obtained exact asymptotic expansions for the dimer model on rectangular $(2M - 1) \times (2N - 1)$ lattices with a single monomer at the boundary under both free and cylindrical boundary conditions. The results revealed that finite-size corrections were significantly influenced by the parity of the lattice sites in both the horizontal and vertical directions. Notably, altering the parity of either dimension results in a change in boundary conditions \cite{Izmailian2005,Izmailian2007}.

It has been shown   \cite{Izmailian2003,Izmailian2019} that the exact asymptotic expansion of the free energy for dimers on an open rectangular $\mathcal{M} \times \mathcal{N}$ lattice takes the following form
\begin{equation}
f = f_{\text{bulk}} + \frac{f_{1s}}{\mathcal{M}} + \frac{f_{2s}}{\mathcal{N}} + f_{\text{corn}}\frac{\ln S}{S} + \frac{f_{0}}{S}+ \sum_{p=1}^{\infty}\frac{f_{p}}{S^{p+1}}\label{freeenergy}.
\end{equation}
Here $S=\mathcal{M} \times \mathcal{N}$ is the area of the lattice. The bulk free energy $f_{\text{bulk}}$ is same for all boundary conditions  \cite{Izmailian2003}:
\begin{equation}
f_\mathrm{bulk}=\frac{1}{2\pi}\int_0^\pi\omega_1(x)dx=\frac{1}{2\pi}\int_0^\pi{\rm arcsinh}{\left( \sin{x}\right)}dx=\frac{G}{\pi},
\label{fbulk}
\end{equation}
where $G$ is the Catalan constant. For the cylindric boundary conditions
the surface free energies are $f_{1s} = - \frac{1}{2}\ln{(1 + \sqrt{2})}=-0.78726...$, $f_{2s} = 0$ and corner free energies $f_{\text{corn}} = 0$ \cite{Izmailian2003}.
The leading finite-size correction term is $f_0$ and the subleading correction terms are $f_p$ for $p = 1,2,3,...$.
This asymptotic form of the free energy is applicable to the Ising model, the spanning-tree model, the Gaussian model, and resistor networks.

The exact asymptotic expansion
of the logarithm of the partition function according the  \cite{Izmailian2002,Izmailian2003} can be written as
\begin{eqnarray}
\ln
Z_{\alpha,\beta}(z,\mathcal{M},\mathcal{N} )&=&\frac{\mathcal{S}}{\pi} \int_{0}^{\pi}\!\!\omega_z(x)~\!{\rm d}x  +
\ln\frac{\theta_{\alpha,\beta}}{\eta}-2\pi\rho\sum_{p=1}^{\infty}
\left(\frac{\pi^2\rho}{\mathcal{S}}\right)^{p}\frac{\Lambda_{2p}}{(2p)!}\,
\frac{{\tt Re}\;{\rm K}_{2p+2}^{{\alpha,\beta}}(i\lambda\rho)}{2p+2}
\label{ExpansionOflnZab}
\end{eqnarray}
Here $\mathcal{S}=\mathcal{M} \times \mathcal{N}$, $\rho = \mathcal{M}/\mathcal{N}$,  $\eta=(\theta_2\theta_3\theta_4/2)^{1/3}$ is the
Dedekind-$\eta$ function; $\theta_{{\alpha,\beta}}$ are elliptic $\theta$-functions with next relations to standard notations are $\theta_{\frac{1}{2},\frac{1}{2}} = \theta_{3}$, $\theta_{\frac{1}{2},0} = \theta_{4}$ and $\theta_{0,\frac{1}{2}} = \theta_{2}$, ${\rm K}_{2p+2}^{{\alpha,\beta}}(i\lambda\rho)$  are Kronecker's double series \cite{Izmailian2002, Weil}, for isotropic case ($z_v=z_h=z=1$) $\int_{0}^{\pi}\!\!\omega_z(x)~\!{\rm d}x = 2 G$ and $G = 0.915966\dots$ is  Catalan's constant.

The differential operators $\Lambda_{2p}$ are defined through the coefficients $\lambda_{2p}$ from the Taylor expansion of the lattice dispersion relation $\omega_0(k)$:
\begin{equation}
\omega_0(k)=k\left(\lambda+\sum_{p=1}^{\infty}
\frac{\lambda_{2p}}{(2p)!}\;k^{2p}\right)
\label{SpectralFunctionExpansion}
\end{equation}
where $\lambda=1$, $\lambda_2=-2/3$, $\lambda_4=4$, etc.

The connections between the differential operators $\Lambda_{2p}$ and the coefficients $\lambda_{2p}$ as detailed in  \cite{Izmailian2002,Izmailian2019}.

Using the results from \cite{Izmailian2003} and by plugging the expression from Equation (\ref{ExpansionOflnZab}) into Equations (\ref{cyl1}), (\ref{cyl2}), and (\ref{cyl3}), we can obtain $f_p(\rho)$  for each of the three types of cylinders.

For dimers on $2M \times 2N$ cylinder one has:
\begin{eqnarray}
f_0(\rho)&=&-\frac{1}{3}\ln{\frac{2\theta_3^2(\rho)}{\theta_2(\rho)\theta_4(\rho)}} = -\ln{\frac{\theta_3(\rho)}{\eta(\rho)}},
\label{f01}\\
f_p(\rho)&=&\frac{2^{-2p-1}{\pi}^{2p+1}\rho^{p+1}}{(2p)!(p+1)}\Lambda_{2p}\,
{\tt Re}\;{\rm K}_{p+1}^{\frac{1}{2},\frac{1}{2}}(2i\lambda\rho),
\label{fp1}
\end{eqnarray}
with
\begin{equation}
S=2M \times N \qquad \mbox{and}  \qquad  \rho = \frac{2M}{N}. \label{Srho1}
\end{equation}

For dimers on $(2M - 1) \times 2N$ cylinder it has been obtained:
\begin{eqnarray}
f_0(\rho)&=&-\frac{1}{3}\ln{\frac{2\theta_4^2(\rho)}{\theta_2(\rho)\theta_3(\rho)}} = -\ln{\frac{\theta_4(\rho)}{\eta(\rho)}},
\label{f02}\\
f_p(\rho)&=&\frac{2^{-2p-1}{\pi}^{2p+1}\rho^{p+1}}{(2p)!(p+1)}\Lambda_{2p}\,
{\tt Re}\;{\rm K}_{p+1}^{\frac{1}{2},0}(2i\lambda\rho),
\label{fp2}
\end{eqnarray}
with
\begin{equation}
S=2M \times N \qquad \mbox{and}  \qquad  \rho = \frac{2M}{N}. \label{Srho2}
\end{equation}

For dimers on $2M \times (2N - 1)$ cylinder it has been received:
\begin{eqnarray}
f_0(\rho)&=&-\frac{1}{6}\ln{\frac{2\theta_2^2(\rho)}{\theta_3(\rho)\theta_4(\rho)}} = -\frac{1}{2}\ln{\frac{\theta_2(\rho)}{\eta(\rho)}},
\label{f03}\\
f_p(\rho)&=&\frac{{\pi}^{2p+1}\rho^{p+1}}{(2p)!(2p+2)}\Lambda_{2p}\,
{\tt Re}\;{\rm K}_{2p+2}^{0,\frac{1}{2}}(2i\lambda\rho),
\label{fp3}
\end{eqnarray}
with
\begin{equation}
S=2M \times 2N \qquad \mbox{and}  \qquad  \rho = \frac{M}{N}. \label{Srho3}
\end{equation}

The expressions for $\Lambda_{2p}$ and the
expressions for $K_{2p+2}^{\frac{1}{2},\frac{1}{2}}(\rho)$ and  $K_{2p+2}^{\frac{1}{2},0}(\rho)$ are given in \cite{Izmailian2019,Izmailian2023}. The expressions for $K_{2p+2}^{0, \frac{1}{2}}(\rho)$ we can obtain
from relations between Kronecker's double series $K_{2p}^{1/2, 1/2}(\tau)$, $K_{2p}^{1/2, 0}(\tau)$ and $K_{2p}^{0, 1/2}(\tau)$  under the transformations $\theta_2 \leftrightarrow \theta_4$ , $\theta_3 \leftrightarrow \theta_4$  and $\theta_2 \leftrightarrow \theta_3$,
received in  Appendix B and C from  \cite{Izmailian2023}. The results for $K_{2p+2}^{0, \frac{1}{2}}(\rho)$ are listed in Supplementary Materials.
Using these expressions for  $\Lambda_{2p}$  and  Kronecker's double series we can express the subleading correction terms $f_p(\rho)$ in the asymptotic expansion of the free energy for the dimer model on the  cylinder in all three cases $2M \times 2N$, $(2M - 1) \times 2N$ and  $2M \times (2N - 1)$ for any value of $p$ in terms of the elliptic theta functions ${\theta}_2, {\theta}_3, {\theta}_4$ and the elliptic integral of the second kind $E$.
In particular in this paper we have calculated the subleading correction terms $f_p$ in terms of the elliptic functions the elliptic integral of the second kind up to $p=17$. Due to very large expressions for $f_p(\rho)$ for $p > 3$ we have not listed those expressions in Appendix\ \ref{fexpression1}, \ref{fexpression2} and \ \ref{fexpression3} for dimers on $2M \times 2N$, $(2M - 1) \times 2N$ and  $2M \times (2N - 1)$ cylinders, respectively.

\begin{figure}[htt!]
  	\includegraphics[width=130mm]{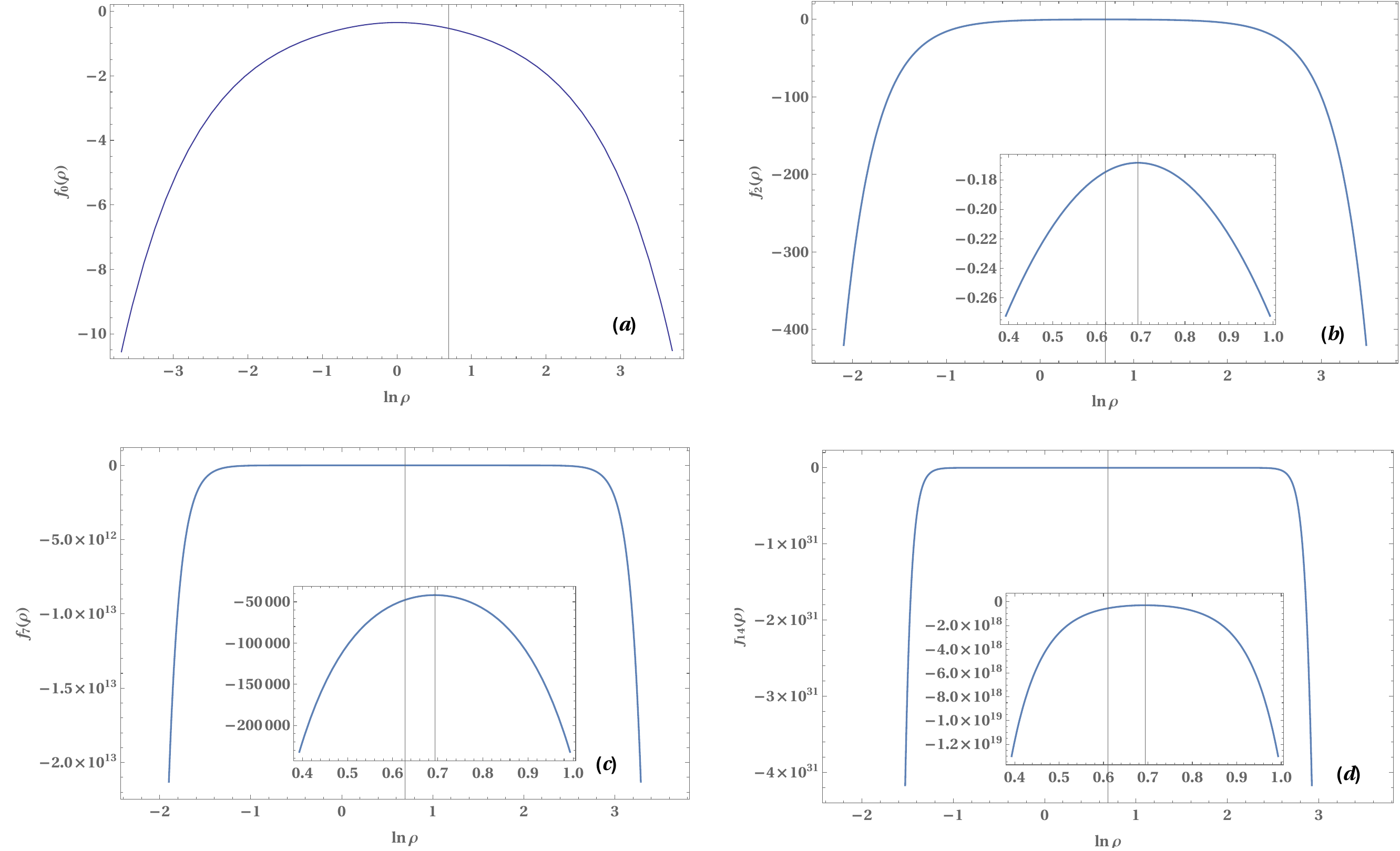}
	\caption{The behavior of the correction
		terms $f_p(\rho)$ of dimers on $2M \times 2N$ cylinder for (a) $p=0$, (b) $p=2$, (c) $p=7$, {and} (d) $p=14$.}
	\label{fp_1}
\end{figure}

\begin{figure}[htt!]
  	\includegraphics[width=130mm]{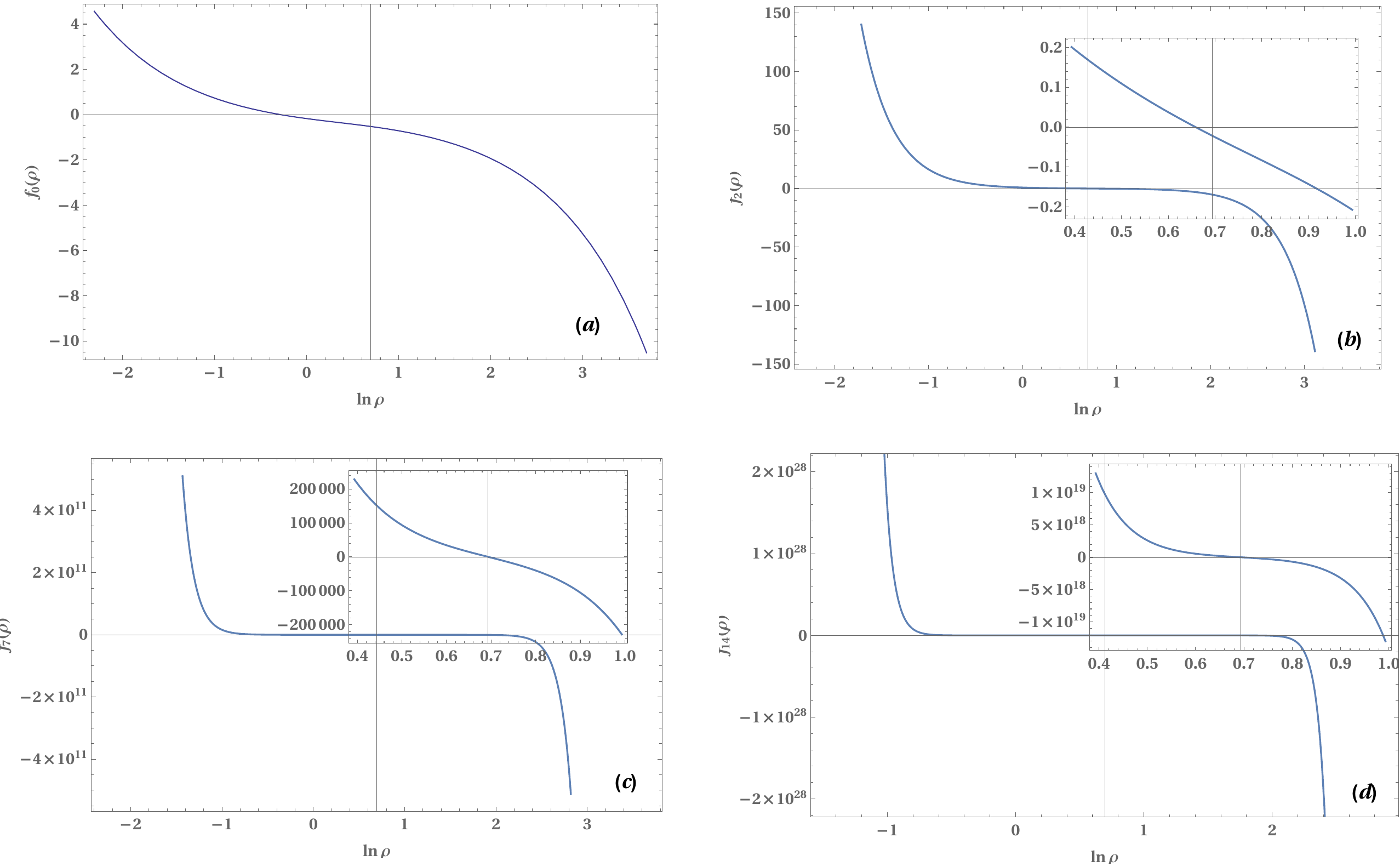}
	\caption{The behavior of the correction
		terms $f_p(\rho)$ of dimers on $(2M - 1) \times 2N$ cylinder for (a) $p=0$, (b) $p=2$, (c) $p=7$, {and} (d) $p=14$.}
	\label{fp_2}
\end{figure}

\begin{figure}[htt!]
  	\includegraphics[width=130mm]{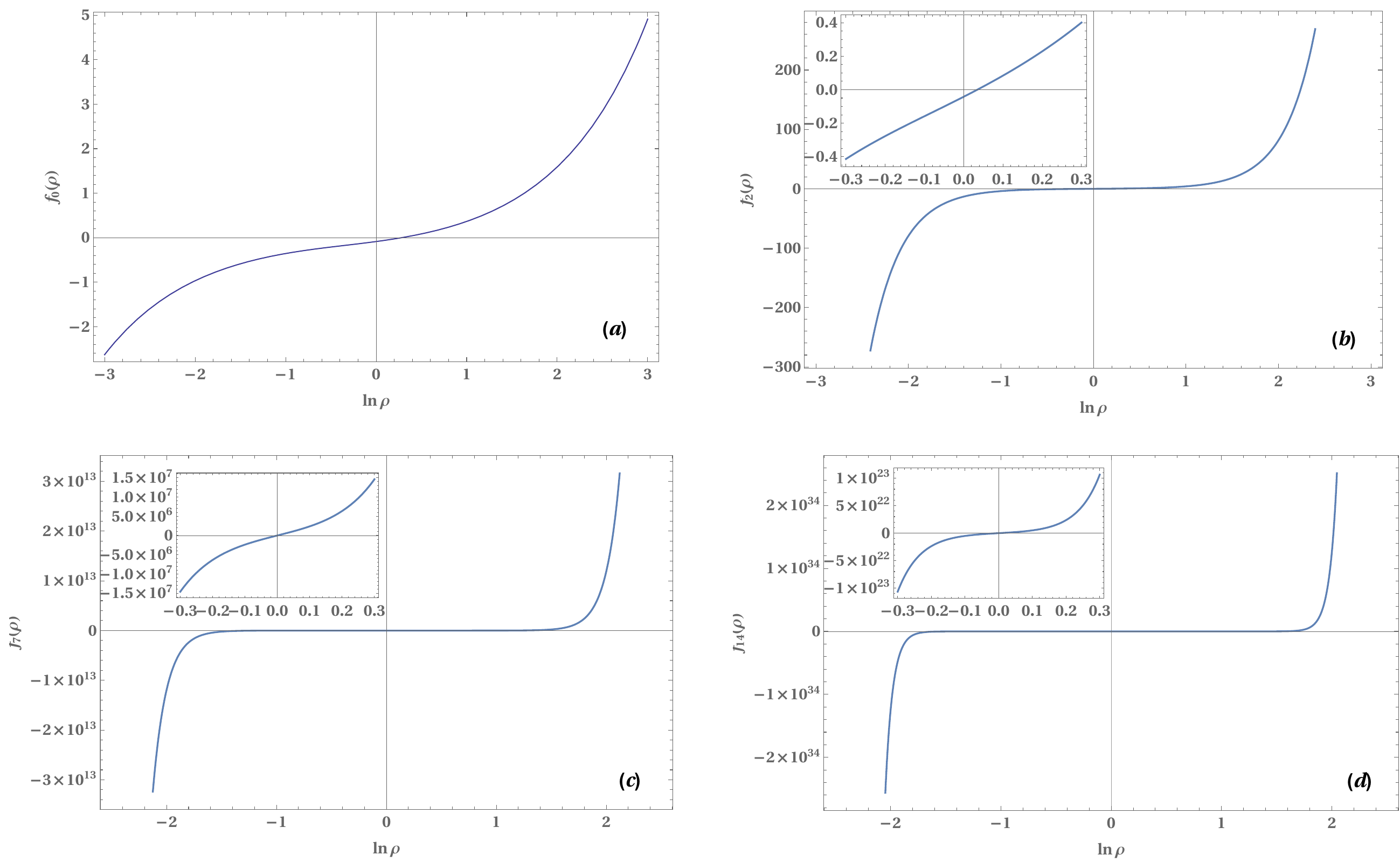}
	\caption{The behavior of the correction
		terms $f_p(\rho)$ of dimers on $2M \times (2N - 1)$ cylinder for (a) $p=0$, (b) $p=2$, (c) $p=7$, {and} (d) $p=14$.}
	\label{fp_3}
\end{figure}

In Fig.\ \ref{fp_1}, \ref{fp_2} and \ref{fp_3} we plot the behavior of the subleading correction terms $f_p(\rho)$ (a) $p=0$, (b) $p=2$, (c) $p=7$, {and} (d) $p=14$ as a function of the aspect ratio $\rho$
for dimers on $2M \times 2N$, $(2M - 1) \times 2N$ and  $2M \times (2N - 1)$ cylinders, respectively.
For dimers on $2M \times 2N$ and $(2M - 1) \times 2N$ cylinders exact numerical values of $f_p(\rho)$, with the aspect ratio $ \rho = 1, 2, 4$ for $p =1, 2, 3,...,17$ are given in Table\ \ref{tabfp1} and \ref{tabfp2}, respectively.
For dimers on $2M \times (2N - 1)$ cylinder exact numerical values of $f_p(\rho)$, with the aspect ratio $ \rho = 1/2, 1, 2$ for $p =1, 2, 3,...,17$ are given in Table\ \ref{tabfp3}.

In Supplementary Materials we have presented the expressions of subleading correction terms $f_p(\rho)$ in terms of gamma function  for the aspect ratio $ \rho = 1/2, 1, 2, 4$ and $p =1, 2, 3,...,8$ obtained using the relation between elliptic theta functions, elliptic integral and gamma function (see Appendix \ \ref{ThetaToGamma}). The expressions for $p > 8$ are so large that they are therefore not given.

\begin{figure}[htt!]
  	\includegraphics[width=130mm]{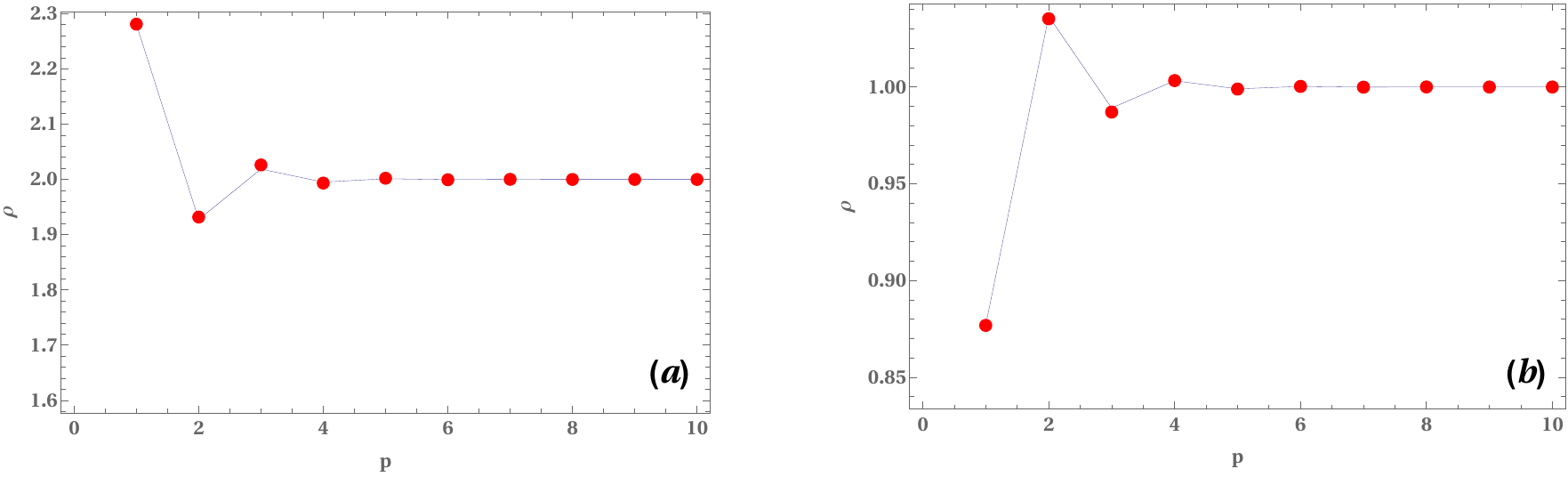}
	\caption{The behavior of roots of the correction
		terms $f_p(\rho)$, as a function of $p$. The dots represent our exact results. For dimers on $(2M - 1) \times 2N$ cylinder (a) the solid line is given by $(-1)^{p} a \, b^p + 2$, with $a=-1.08577$ and $b=0.25831$. For dimers on $2M \times (2N - 1)$ cylinder (b) the solid line is given by $(-1)^{p+1} a \, b^p + 1$, with $a=-0.41405$ and $b=0.29709$.}
	\label{roots}
\end{figure}

We noticed that the roots coefficients  $f_p(\rho)$ as a function of the aspect ratio $\rho$ for increasing
$p$ exponentially tends to $2$ for dimers on $(2M - 1) \times 2N$ cylinder and to $1$ for dimers on $2M \times (2N - 1)$ cylinder. These results are shown in the Fig.\ \ref{roots}.
We also found that the maximum of the function $f_p(\rho)$ for dimers on $2M \times 2N$ cylinder is reached at $\rho = 2$, which is also seen in Fig.\ \ref{fp_1}.

\begin{table}[ht!]
	\caption{Coefficients $f_p$ in the asymptotic expansion of the free energy for model of dimers on $2M \times 2N$ cylinder, for the aspect ratio 1, 2, 4.}
	\label{tabfp1}
	\begin{center}
		\begin{tabular}{| c | c | c | c |}
			\hline
			$p$ & $f_{p}(1)$ & $f_{p}(2)$ & $f_{p}(4)$ \\\hline
			0 & -0.346573590280... & -0.527330191080... & -1.04720452587... \\\hline
			1 & -0.320716568887... & -0.188074112577... & -0.320716568887... \\\hline
			2 & -0.810301005787... & -0.168437484866... & -0.810301005787... \\\hline
			3 & -6.27315437483... & -0.822513387616... & -6.27315437483... \\\hline
			4 & $-9.80532584302...\times 10^1$ & $-0.601821634381...\times 10^1$ & $-9.80532584302...\times 10^1$ \\\hline
			5 & $-0.254149909928...\times 10^4$ & $-0.798824718897...\times 10^2$ & $-0.254149909928...\times 10^4$ \\\hline
			6 &$ -9.85769596813...\times 10^4$ & $-0.153686749144...\times 10^4$ & $-9.85769596813...\times 10^4$ \\\hline
			7 & $-5.34679715780...\times 10^6$ & $-4.18016773945...\times 10^4$ & $-5.34679715780...\times 10^6$ \\\hline
			8 & $-3.86387617973...\times 10^8$ & $-1.50895329542...\times 10^6$ & $-3.86387617973...\times 10^8$ \\\hline
			9 & $-3.58814194064...\times 10^{10}$ & $-7.00865228737...\times 10^7$ & $-3.58814194064...\times 10^{10}$ \\\hline
			10 &$-4.16355939979...\times 10^{12}$ & $-4.06586520894...\times 10^9$ & $-4.16355939979...\times 10^{12}$ \\\hline
			11 & $-5.90325407711...\times 10^{14}$ & $-2.88247381999...\times 10^{11}$ & $-5.90325407711...\times 10^{14}$ \\\hline
			12 & $-1.00417540827...\times 10^{17}$ & $-2.45159289451...\times 10^{13}$ & $-1.00417540827...\times 10^{17}$ \\\hline
			13 & $-2.01840674021...\times 10^{19}$ & $-2.46387777220...\times 10^{15}$ & $-2.01840674021...\times 10^{19}$ \\\hline
			14 & $-4.73259659515...\times 10^{21}$ & $-2.88854681755...\times 10^{17}$ & $-4.73259659515...\times 10^{21}$ \\\hline
			15 & $-1.28024730925...\times 10^{24}$ & $-3.90700512651...\times 10^{19}$ & $-1.28024730925...\times 10^{24}$ \\\hline
			16 & $-3.95771049508...\times 10^{26}$ & $-6.03898676038...\times 10^{21}$ & $-3.95771049508...\times 10^{26}$ \\\hline
			17 & $-1.38650652641...\times 10^{29}$ & $-1.05782054241...\times 10^{24}$ & $-1.38650652641...\times 10^{29}$ \\\hline			
		\end{tabular}
	\end{center}
\end{table}

\begin{table}[ht!]
	\caption{Coefficients $f_p$ in the asymptotic expansion of the free energy for model of dimers on $(2M - 1) \times 2N$ cylinder, for the aspect ratio 1, 2, and 4.}
	\label{tabfp2}
	\begin{center}
		\begin{tabular}{| c | c | c | c |}
			\hline
			$p$ & $f_{p}(1)$ & $f_{p}(2)$ & $f_{p}(4)$ \\\hline
			0 & -0.173286795140... & -0.519860385420... & -1.04719057650... \\\hline
			1 & 0.344225832959... & 0.0470185281444... & -0.282111168866... \\\hline
			2 & 0.805037334385... & -0.0210546856082... & -0.757968681896... \\\hline
			3 & 6.27827802614... & 0.0409892105392... & -6.25219341661... \\\hline
			4 & $9.80469113511...\times 10^1$ & $-0.101553266149...$ &
$-9.79163137594...\times 10^1$ \\\hline
			5 & $0.254151503399...\times 10^4$ & $0.509910660391...$ &
$-0.253992195934...\times 10^4$ \\\hline
			6 &$ 0.985769098723...\times 10^5$ & $-3.18777342933...$
& $-0.985635369519...\times 10^5$ \\\hline
			7 & $5.34679739918...\times 10^6$ & $0.308966466082...\times 10^2$ & $-5.34665993573...\times 10^6$ \\\hline
			8 & $3.86387616558...\times 10^8$ & $-0.362149183062...\times 10^3$ & $-3.86384753819...\times 10^8$ \\\hline
			9 & $3.58814194176...\times 10^{10}$ & $0.577022853154...\times 10^4$ & $-3.58813453543...\times 10^{10}$ \\\hline
			10 &$4.16355939969...\times 10^{12}$ & $-1.08826842309...\times 10^5$ & $-4.16355741264...\times 10^{12}$ \\\hline
			11 & $5.90325407712...\times 10^{14}$ & $2.58698242660...\times 10^6$ & $-5.90325340194...\times 10^{14}$ \\\hline
			12 & $1.00417540827...\times 10^{17}$ & $-7.14982909181...\times 10^7$ & $-1.00417537816...\times 10^{17}$ \\\hline
			13 & $2.01840674021...\times 10^{19}$ & $2.37560308928...\times 10^9$ & $-2.01840672489...\times 10^{19}$ \\\hline
			14 & $4.73259659515...\times 10^{21}$ & $-9.03636358888...\times 10^{10}$ & $-4.73259658638...\times 10^{21}$ \\\hline
			15 & $1.28024730925...\times 10^{24}$ & $4.00370504841...\times 10^{12}$ & $-1.28024730866...\times 10^{24}$ \\\hline
			16 & $3.95771049508...\times 10^{26}$ & $-2.00398753867...\times 10^{14}$ & $-3.95771049461...\times 10^{26}$ \\\hline
			17 & $1.38650652641...\times 10^{29}$ & $1.14262763047...\times 10^{16}$ & $-1.38650652637...\times 10^{29}$ \\\hline			
		\end{tabular}
	\end{center}
\end{table}

\begin{table}[ht!]
	\caption{Coefficients $f_p$ in the asymptotic expansion of the free energy for model of dimers on $2M \times (2N - 1)$ cylinder, for the aspect ratio $\frac{1}{2}$, 1, and 2.}
	\label{tabfp3}
	\begin{center}
		\begin{tabular}{| c | c | c | c |}
			\hline
			$p$ & $f_{p}(\frac{1}{2})$ & $f_{p}(1)$ & $f_{p}(2)$ \\\hline
			0 & -0.259930192710... & -0.086643397570... & 0.177021697970... \\\hline
			1 & -0.282111168866... & 0.0470185281444... & 0.344225832959... \\\hline
			2 & -1.51593736379... & -0.0421093712164... & 1.61007466877... \\\hline
			3 & $-2.50087736664...\times 10^1$ & 0.163956842157... & $0.251131121046...\times 10^2 $ \\\hline
			4 & $-0.783330510075...\times 10^3$ & $-0.812426129192...$ &
$0.784375290809...\times 10^3$ \\\hline
			5 & $-0.406387513494...\times 10^5$ & $0.815857056626...\times 10^1$ &
$0.406642405438...\times 10^5$ \\\hline
			6 &$ -0.315403318246...\times 10^7$ & $-0.102008749739...\times 10^3$ & $3.15446111591...\times 10^6$ \\\hline
			7 & $-0.342186235887...\times 10^9$ & $0.197738538293...\times 10^4$ & $3.42195033547...\times 10^8$ \\\hline
			8 & $-0.494572484888...\times 10^{11}$ & $-0.463550954319...\times 10^5$ & $4.94576149194...\times 10^{10}$ \\\hline
			9 & $-9.18562441070...\times 10^{12}$ & $1.47717850408...\times 10^6$ & $9.18564337091...\times 10^{12}$ \\\hline
			10 &$-2.13174139527...\times 10^{15}$ & $-5.57193432624...\times 10^7$ & $2.13174241264...\times 10^{15}$ \\\hline
			11 & $-6.04493148359...\times 10^{17}$ & $2.64907000484...\times 10^9$ & $6.04493217497...\times 10^{17}$ \\\hline
			12 & $-2.05655117448...\times 10^{20}$ & $-1.46428499800...\times 10^{11}$ & $2.05655123614...\times 10^{20}$ \\\hline
			13 & $-8.26739394514...\times 10^{22}$ & $9.73047025371...\times 10^{12}$ & $8.26739400791...\times 10^{22}$ \\\hline
			14 & $-3.87694312357...\times 10^{25}$ & $-7.40258905201...\times 10^{14}$ & $3.87694313074...\times 10^{25}$ \\\hline
			15 & $-2.09755719051...\times 10^{28}$ & $6.55967035132...\times 10^{16}$ & $-2.09755719148...\times 10^{28}$ \\\hline
			16 & $-1.29686257488...\times 10^{31}$ & $-6.56666636671...\times 10^{18}$ & $1.29686257503...\times 10^{31}$ \\\hline
			17 & $-9.08660917119...\times 10^{33}$ & $7.48832443907...\times 10^{20}$ & $9.08660917146...\times 10^{33}$ \\\hline			
		\end{tabular}
	\end{center}
\end{table}

\section{Dimer model on infinitely long strip and cylinder}

Using Kronecker's functions asymptotic form (see for example  \cite{Izmailian2019,Izmailian2023}) when
$\rho = \mathcal{M}/\mathcal{N} \to \infty$  (i.e. ${\mathcal M} \to \infty$) for fixed ${\mathcal N}$ from Eq. (\ref{freeenergy}) one obtains the free energy
expansion for infinitely long cylinder of circumference $2 {\mathcal N}$ with periodic boundary conditions
\begin{eqnarray}
\lim_{{\mathcal M} \to \infty}\frac{F}{{\mathcal M}}&=&{\mathcal N} f_{bulk}-\frac{\pi}{24 {\mathcal N}}
+\sum_{p=1}^{\infty}\frac{f_p^\mathrm{cyl}}{{\mathcal N}^{2p+1}}, \label{cylinder}
\end{eqnarray}
where $f_p^\mathrm{cyl}$ is given by
\begin{eqnarray}
f_p^\mathrm{cyl} = \lim_{\rho \to \infty}\rho^{-p-1}f_p(\rho). \label{fpfcylinder}
\end{eqnarray}

In the limit $ \rho \to 0$ (i.e. ${\mathcal N} \to \infty$) for fixed ${\mathcal M}$ we obtain the expansion of free energy of infinitely long strip with free boundary condition of the width ${\mathcal M}$
\begin{eqnarray}
\lim_{{\mathcal{N}} \to \infty}\frac{F}{{\mathcal N}}&=&{\mathcal M} f_{bulk}+f_{1s}
+\frac{\pi}{24{\mathcal M}}+\sum_{p=1}^{\infty}\frac{f_p^\mathrm{str}}{{\mathcal M}^{2p+1}},
\label{strip}
\end{eqnarray}
where $f_p^\mathrm{str}$ is given by
\begin{eqnarray}
f_p^\mathrm{str} = \lim_{\rho \to 0}\rho^{p+1}f_p(\rho). \label{fpfstrip}
\end{eqnarray}

By performing the corresponding limit transitions as in \cite{Izmailian2019,Izmailian2023,Izmailian2002a}) for the coefficients, we obtain:

for dimers on $2M \times 2N$ cylinder  with $\rho = 2M/N$:
\begin{eqnarray}
f_p^\mathrm{cyl}=\frac{2^{-2p-1}\pi^{2p+1}\lambda_{2p}B_{2p+2}^{1/2}}{(2p)!(p+1)},\label{fpcylinder1}
\end{eqnarray}
\begin{eqnarray}
f_p^\mathrm{str}=\frac{2\pi^{2p+1}\lambda_{2p}B_{2p+2}^{1/2}}{(2p)!(p+1)},\label{fpstrip1}
\end{eqnarray}

\begin{table}[ht]
	\caption{Coefficients $f_p$ for dimers on $2M \times 2N$ cylinder in the asymptotic expansion of the free energy for the infinite strip ($f_{p}^\mathrm{str}$) and for the infinite cylinder ($f_{p}^\mathrm{cyl}$).}
	\label{tabfsc1}
	\begin{center}
		\begin{tabular}{| c | c | c |}
			\hline
			$p$ & $f_{p}^\mathrm{str}$ & $f_{p}^\mathrm{cyl}$  \\\hline
			0 & -0.523598775598... & -0.130899693900... \\\hline
			1 & -0.301449912170... & -0.0188406195106... \\\hline
			2 & -0.784276639248... & -0.0122543224882... \\\hline
			3 & -6.26293856529... & -0.0244646037707... \\\hline
			4 & $ -9.79850293217...\times 10^{1} $ & -0.0956885051969 \\\hline
			5 & $ -0.254071064704...\times 10^{4} $ & -0.620290685311 \\\hline
			6 & $ -0.985702483672...\times 10^{5} $ & $ -0.601625051070...\times 10^{1} $ \\\hline
			7 & $ -5.34672854693...\times 10^6 $ & $ -0.815846030721...\times 10^2 $ \\\hline
			8 & $ -3.86386185898...\times 10^8 $ & $ -0.147394632682...\times 10^4 $ \\\hline
			9 & $ -3.58813823804...\times 10^{10} $ & $ -0.342191528133...\times 10^{5} $ \\\hline
			10 & $ -4.16355840622...\times 10^{12} $ & $ -0.992669679216...\times 10^{6} $ \\\hline
			11 & $ -5.90325373952...\times 10^{14} $ & $ -3.51861342163...\times 10^7 $ \\\hline
			12 & $ -1.00417539322...\times 10^{17} $ & $ -1.49633794012\times 10^9 $ \\\hline
			13 & $ -2.01840673255...\times 10^{19} $ & $ -7.51915101912...\times 10^{10} $ \\\hline
			14 & $ -4.73259659077...\times 10^{21} $ & $ -4.40757404153...\times 10^{12} $ \\\hline
			15 & $ -1.28024730896...\times 10^{24} $ & $ -2.98080804981...\times 10^{14} $ \\\hline
			16 & $ -3.95771049485...\times 10^{26} $ & $ -2.30369070478...\times 10^{16} $ \\\hline
			17 & $ -1.38650652639...\times 10^{29} $ & $ -2.01763254355...\times 10^{18} $ \\\hline			
		\end{tabular}
	\end{center}
\end{table}

for dimers on $(2M -1) \times 2N$ cylinder  with $\rho = 2M/N$:
\begin{eqnarray}
f_p^\mathrm{cyl}=\frac{2^{-2p-1}\pi^{2p+1}\lambda_{2p}B_{2p+2}^{1/2}}{(2p)!(p+1)},\label{fpcylinder2}
\end{eqnarray}
\begin{eqnarray}
f_p^\mathrm{str}=\frac{2\pi^{2p+1}\lambda_{2p}B_{2p+2}}{(2p)!(p+1)},\label{fpstrip2}
\end{eqnarray}

\begin{table}[ht]
	\caption{Coefficients $f_p$ for dimers on $(2M - 1) \times 2N$ cylinder in the asymptotic expansion of the free energy for the infinite strip ($f_{p}^\mathrm{str}$) and for the infinite cylinder ($f_{p}^\mathrm{cyl}$).}
	\label{tabfsc2}
	\begin{center}
		\begin{tabular}{| c | c | c |}
			\hline
			$p$ & $f_{p}^\mathrm{str}$ & $f_{p}^\mathrm{cyl}$  \\\hline
			0 & 1.04719755120... & -0.130899693900... \\\hline
			1 & 0.344514185337... & -0.0188406195106... \\\hline
			2 & 0.809575885675... & -0.0122543224882... \\\hline
			3 & 6.31225304218... & -0.0244646037707... \\\hline
			4 & $ 9.81767808468...\times 10^{1} $ & -0.0956885051969 \\\hline
			5 & $ 0.254195183445...\times 10^{4} $ & -0.620290685311 \\\hline
			6 & $ 0.985822823373...\times 10^{5} $ & $ -0.601625051070...\times 10^{1} $ \\\hline
			7 & $ 5.34689172112...\times 10^6 $ & $ -0.815846030721...\times 10^2 $ \\\hline
			8 & $ 3.86389133814...\times 10^8 $ & $ -0.147394632682...\times 10^4 $ \\\hline
			9 & $ 3.58814508188...\times 10^{10} $ & $ -0.342191528133...\times 10^{5} $ \\\hline
			10 & $ 4.16356039156...\times 10^{12} $ & $ -0.992669679216...\times 10^{6} $ \\\hline
			11 & $ 5.90325444325...\times 10^{14} $ & $ -3.51861342163...\times 10^7 $ \\\hline
			12 & $ 1.00417542314...\times 10^{17} $ & $ -1.49633794012\times 10^9 $ \\\hline
			13 & $ 2.01840674759...\times 10^{19} $ & $ -7.51915101912...\times 10^{10} $ \\\hline
			14 & $ 4.73259659958...\times 10^{21} $ & $ -4.40757404153...\times 10^{12} $ \\\hline
			15 & $ 1.28024730955...\times 10^{24} $ & $ -2.98080804981...\times 10^{14} $ \\\hline
			16 & $ 3.95771049531...\times 10^{26} $ & $ -2.30369070478...\times 10^{16} $ \\\hline
			17 & $ 1.38650652643...\times 10^{29} $ & $ -2.01763254355...\times 10^{18} $ \\\hline			
		\end{tabular}
	\end{center}
\end{table}

for dimers on $2M \times (2N - 1)$ cylinder  with $\rho = M/N$:
\begin{eqnarray}
f_p^\mathrm{cyl}=\frac{\pi^{2p+1}\lambda_{2p}B_{2p+2}}{(2p)!(2p+2)},\label{fpcylinder3}
\end{eqnarray}
\begin{eqnarray}
f_p^\mathrm{str}=\frac{\pi^{2p+1}\lambda_{2p}B_{2p+2}^{1/2}}{(2p)!2p+2)}.\label{fpstrip3}
\end{eqnarray}

\begin{table}[ht]
	\caption{Coefficients $f_p$ for dimers on $2M \times (2N -1)$ cylinder in the asymptotic expansion of the free energy for the infinite strip ($f_{p}^\mathrm{str}$) and for the infinite cylinder ($f_{p}^\mathrm{cyl}$).}
	\label{tabfsc3}
	\begin{center}
		\begin{tabular}{| c | c | c |}
			\hline
			$p$ & $f_{p}^\mathrm{str}$ & $f_{p}^\mathrm{cyl}$  \\\hline
			0 & -0.130899693900... & 0.261799387799... \\\hline
			1 & -0.0753624780424... & 0.0861285463342... \\\hline
			2 & -0.196069159812... & 0.202393971419... \\\hline
			3 & -1.56573464132... & 1.57806326054... \\\hline
			4 & $ -2.44962573304...\times 10^{1} $ & $ 2.45441952117...\times 10^{1} $ \\\hline
			5 & $ -0.635177661759...\times 10^{3} $ & $ 0.635487958614...\times 10^{3} $ \\\hline
			6 & $ -0.246425620918...\times 10^{5} $ & $ 0.246455705843...\times 10^{5} $ \\\hline
			7 & $ -1.33668213673...\times 10^6 $ & $ 1.33672293028...\times 10^6 $ \\\hline
			8 & $ -9.65965464746...\times 10^7 $ & $ 9.65972834534...\times 10^7 $ \\\hline
			9 & $ -8.97034559509...\times 10^{9} $ & $ 8.97036270470...\times 10^9 $ \\\hline
			10 & $ -1.04088960155...\times 10^{12} $ & $ 1.04089009789...\times 10^{12} $ \\\hline
			11 & $ -1.47581343488...\times 10^{14} $ & $ 1.47581361081...\times 10^{14} $ \\\hline
			12 & $ -2.51043848305...\times 10^{16} $ & $ 2.51043855786...\times 10^{16} $ \\\hline
			13 & $ -5.04601683138...\times 10^{18} $ & $ 5.04601686897...\times 10^{18} $ \\\hline
			14 & $ -1.18314914769...\times 10^{21} $ & $ 1.18314914990...\times 10^{21} $ \\\hline
			15 & $ -3.20061827240...\times 10^{23} $ & $ 3.20061827389...\times 10^{23} $ \\\hline
			16 & $ -9.89427623712...\times 10^{25} $ & $ 9.89427623827...\times 10^{25} $ \\\hline
			17 & $ -3.46626631597...\times 10^{28} $ & $ 3.46626631607...\times 10^{28} $ \\\hline			
		\end{tabular}
	\end{center}
\end{table}

We have also used in all three cases the relation of the Bernoulli numbers $B_n\equiv B_n^{0}$ and $B_n^{1/2}=(2^{1-n}-1)B_n$.

Exact numerical values of $f_p^\mathrm{cyl}$ and $f_p^\mathrm{str}$
for dimers on $2M \times 2N$, $(2M -1) \times 2N$ and $(2M -1) \times 2N$ cylinders  are given in Table\ \ref{tabfsc1}, \ref{tabfsc3} and \ref{tabfsc3}, respectively. The exact expressions are given for all three cases in Supplementary Materials.

\section{The ratio of the coefficients $f_p(\rho)$ in the free energy expansion}

Now we consider ratios $r_p$ for the dimer model on cylinder analogous the \cite{Izmailian2019,Izmailian2023}.
The ratio of the coefficients in the free energy expansion $f_p(\rho)$ for $\rho < \rho_{0}$ multiplied by $\rho^{p+1}$ and the coefficients $f_p$  for the strip $f_p^\mathrm{str}$  is written as
\begin{eqnarray}
r_p^\mathrm{str}(\rho)=\frac{\rho^{p+1} f_{p}(\rho)}{f_{p}^\mathrm{str}}
\label{ratiostr}
\end{eqnarray}
and the ratio of the coefficients in the free energy expansion $f_p(\rho)$ for $\rho > \rho_{0}$ multiplied by $\rho^{-p-1}$ and the coefficients $f_p$ for the infinitely long cylinder $f_p^\mathrm{cyl}$
as
\begin{eqnarray}
r_p^\mathrm{cyl}(\rho)=\frac{\rho^{-p-1} f_{p}(\rho)}{f_{p}^\mathrm{cyl}}.
\label{ratiocyl}
\end{eqnarray}

\begin{figure}[htt!]
  	\includegraphics[width=130mm]{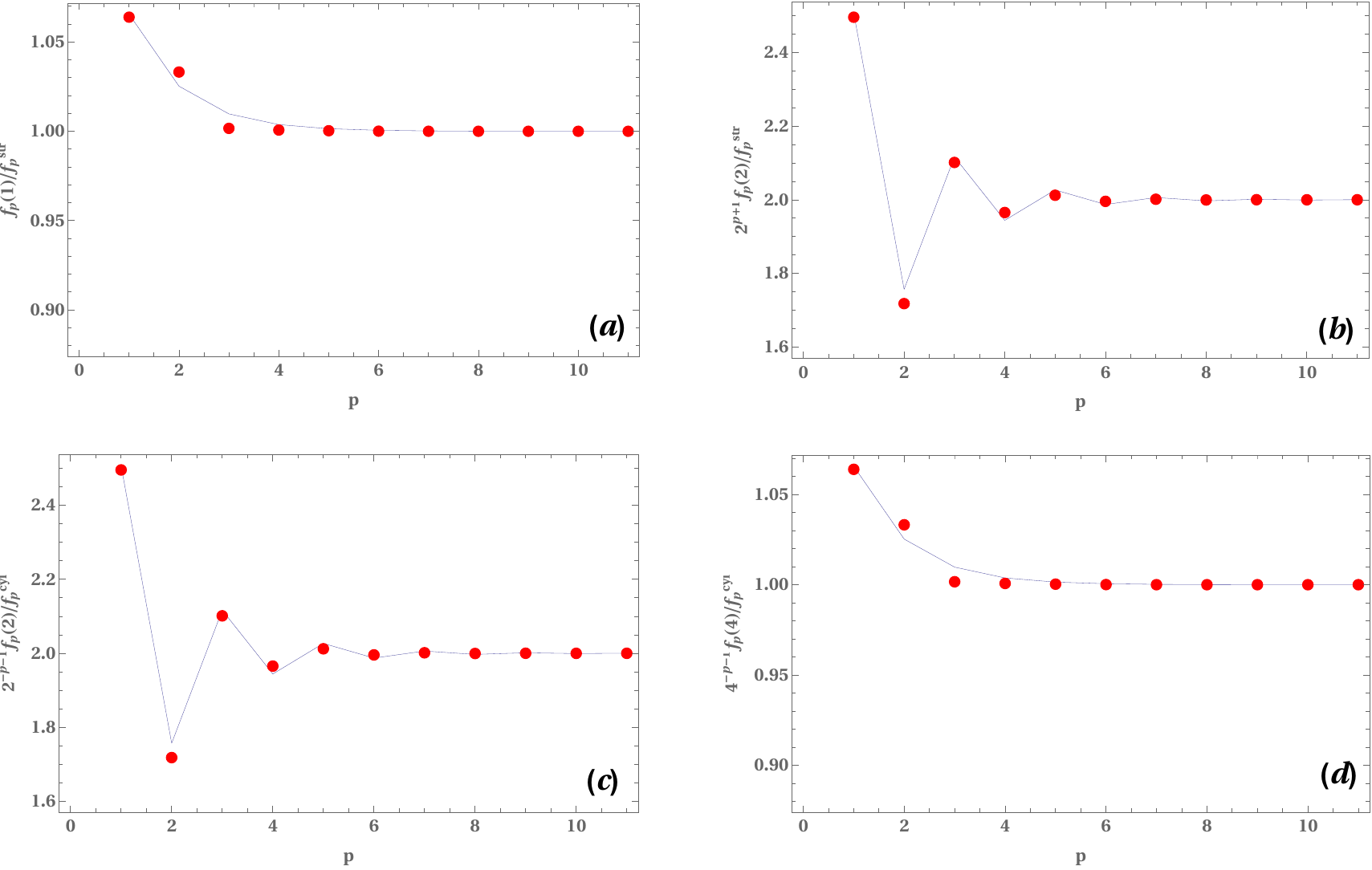}
	\caption{The behavior of ratio of the correction
		terms $f_p(\rho)$, as a function of $p$ for dimers on $2M\times 2N$ cylinder. The dots represent our exact results. For $\rho = 1$ (a) the solid line is given by $a \, b^p + 1$, with $a=0.17017$ and $b=0.38525$. For $\rho = 2$  (b) and (c) the solid line is given by $(-1)^{p} a \, b^p + 2$, with $a=-1.05881$ and $b=0.47912$. For $\rho = 4$  (d) the solid line is given by $a \, b^p + 1$., with $a=0.17017$ and $b=0.38525$. }
	\label{ratio_1}
\end{figure}

\begin{figure}[htt!]
  	\includegraphics[width=130mm]{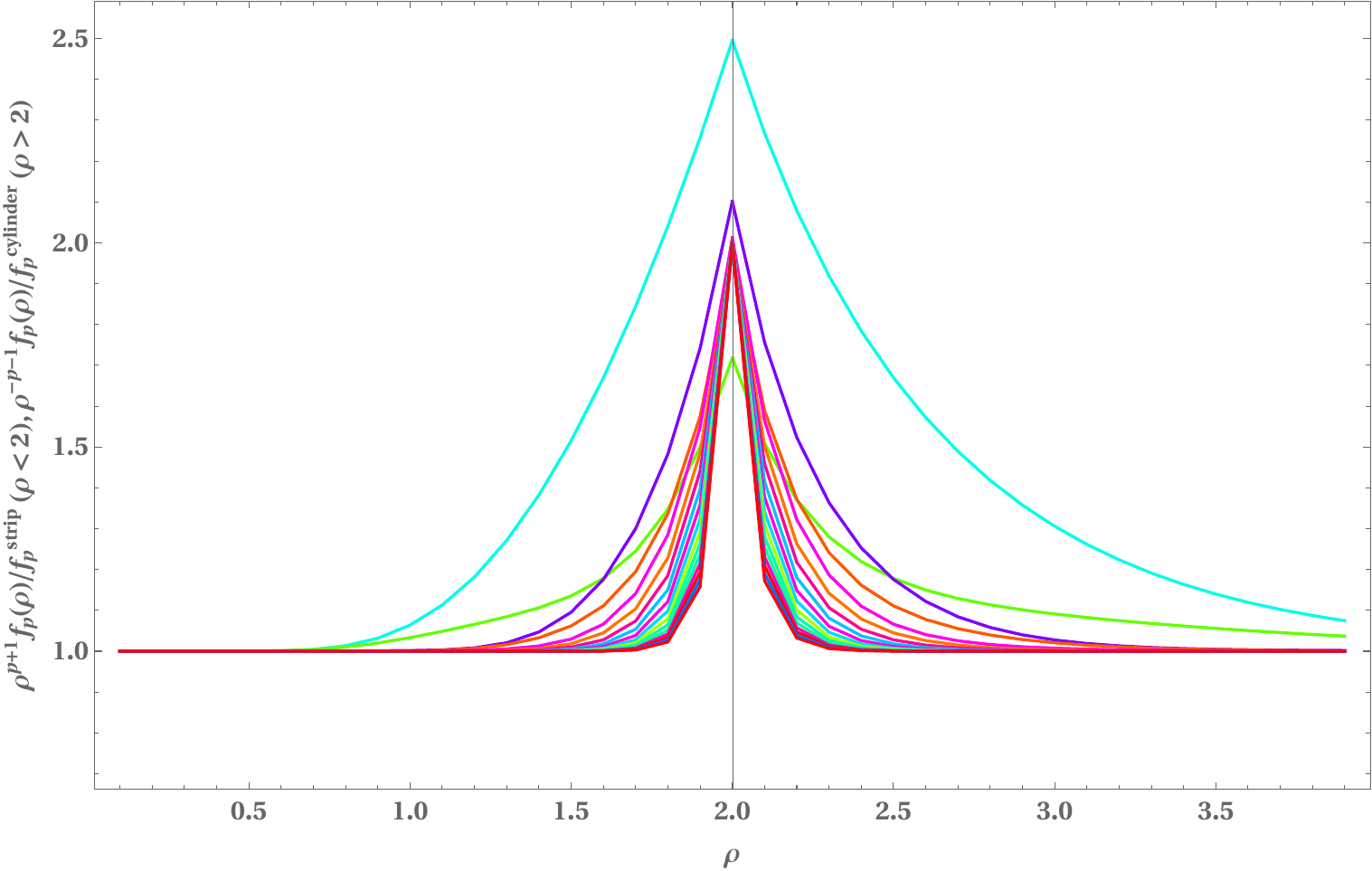}
	\caption{The behavior of ratios of the coefficients $f_p(\rho)$ for dimers on $2M \times 2N$ cylinder in the asymptotic expansion of the free energy  with aspect ratio $\rho < 2$ times $\rho^{p+1}$  to the asymptotic coefficients for the strip $f_p^\mathrm{str}$ and the coefficients $f_p(\rho)$ with aspect ratio $\rho > 2$ times $\rho^{-p-1}$ to the asymptotic coefficients for the cylinder $f_p^\mathrm{cyl}$, as a function of $\rho$ for p =1, 2, 3,...,17.}
	\label{f_rho_1}
\end{figure}

For the dimers on $2M \times 2N$ cylinder the $\rho_{0} = 2$.
We list the values of $r_p^\mathrm{str}(\rho)$ and $r_p^\mathrm{cyl}(\rho)$ for $p=0$ up to $p=17$ in Table\ \ref{tabr1} and plot them against $p$ in  Fig.\ \ref{ratio_1}. The behavior of ratios depending on $\rho$ shown in Fig.\ \ref{f_rho_1}.
One can clearly see that the ratios $r_p^\mathrm{str}(\rho)$ and $r_p^\mathrm{cyl}(\rho)$ exponentially tend to 1 in both cases as $p$ increase and $\rho\neq 2$. From  Table\ \ref{tabr1} and  Figs.\ \ref{ratio_1} and \ \ref{f_rho_1}, one also sees that the ratio  $r_p$      exponentially tends to $2$ for $\rho = 2$ and large values of $p$.
\begin{table}[ht!]
	\caption{Ratios of the coefficients in the asymptotic expansion of the free energy $f_p(\rho)$ for dimers on $2M \times 2N$ cylinder with $\rho = 1$ to the asymptotic coefficients for the strip $f_p^\mathrm{str}$, $f_p(\rho)$ with $\rho = 2$ times $2^{p + 1}$ to the asymptotic coefficients for the strip $f_p^\mathrm{str}$ , $f_p(\rho)$ with $\rho = 2$ times $2^{-p-1}$ to the asymptotic coefficients for the infinite cylinder $f_p^\mathrm{cyl}$ and $f_p(\rho)$ with $\rho = 4$ times $4^{-p-1}$ to the asymptotic coefficients for the infinite cylinder $f_p^\mathrm{cyl}$ as a function of $p$.}
	\label{tabr1}
	\begin{center}
		\begin{tabular}{| c | c | c | c | c |}
			\hline
			$p$ & $r_p^\mathrm{str}(1)=\frac{f_p(1)}{f_{p}^\mathrm{str}}$ & $r_p^\mathrm{str}(2)=\frac{2^{p + 1} f_p(2)}{f_{p}^\mathrm{str}}$ & $r_p^\mathrm{cyl}(2)=\frac{2^{-p-1} f_p(2)}{f_{p}^\mathrm{cyl}}$ & $r_p^\mathrm{cyl}(4)=\frac{4^{-p-1} f_p(4)}{f_{p}^\mathrm{cyl}}$ \\\hline
			0 & 0.661906800458... & 2.01425295725... & 2.01425295725... & 2.00001332064... \\\hline
			1 & 1.06391329352... & 2.49559352960... & 2.49559352960... & 1.06391329352... \\\hline
			2 & 1.03318263638... & 1.71814358798... & 1.71814358798... & 1.03318263638... \\\hline
			3 & 1.00163115276... & 2.10128425573... & 2.10128425573... &  1.00163115276... \\\hline
			4 & 1.00069632177... & 1.96543211075... & 1.96543211075... & 1.00069632177... \\\hline
			5 & 1.00031032744... & 2.01222370871... & 2.01222370871... & 1.00031032744... \\\hline
			6 & 1.00006808661... & 1.99572428967... & 1.99572428967... & 1.00006808661... \\\hline
			7 & 1.00001283231... & 2.00145365882... & 2.00145365882... & 1.00001283231... \\\hline
			8 & 1.00000370633... & 1.99951270375... & 1.99951270375... & 1.00000370633... \\\hline
			9 & 1.00000103190... & 2.00016260973... & 2.00016260973... & 1.00000103190... \\\hline
			10 & 1.00000023864... & 1.99994599222... & 1.99994599222... & 1.00000023864... \\\hline
			11 & 1.00000005719... & 2.00001783553... & 2.00001783553... & 1.00000005719... \\\hline
			12 & 1.00000001499... & 1.99999413723... & 1.99999413723... & 1.00000001499... \\\hline
			13 & 1.00000000380... & 2.00000192075... & 2.00000192075... & 1.00000000380... \\\hline
			14 & 1.00000000093... & 1.99999937248... & 1.99999937248... & 1.00000000093... \\\hline
			15 & 1.00000000023... & 2.00000020449... & 2.00000020449... & 1.00000000023... \\\hline
			16 & 1.00000000006... & 1.99999993351... & 1.99999993351... & 1.00000000006... \\\hline
			17 & 1.00000000001... & 2.00000002157... & 2.00000002157... & 1.00000000001... \\\hline
		\end{tabular}
	\end{center}
\end{table}

\begin{figure}[htt!]
  	\includegraphics[width=130mm]{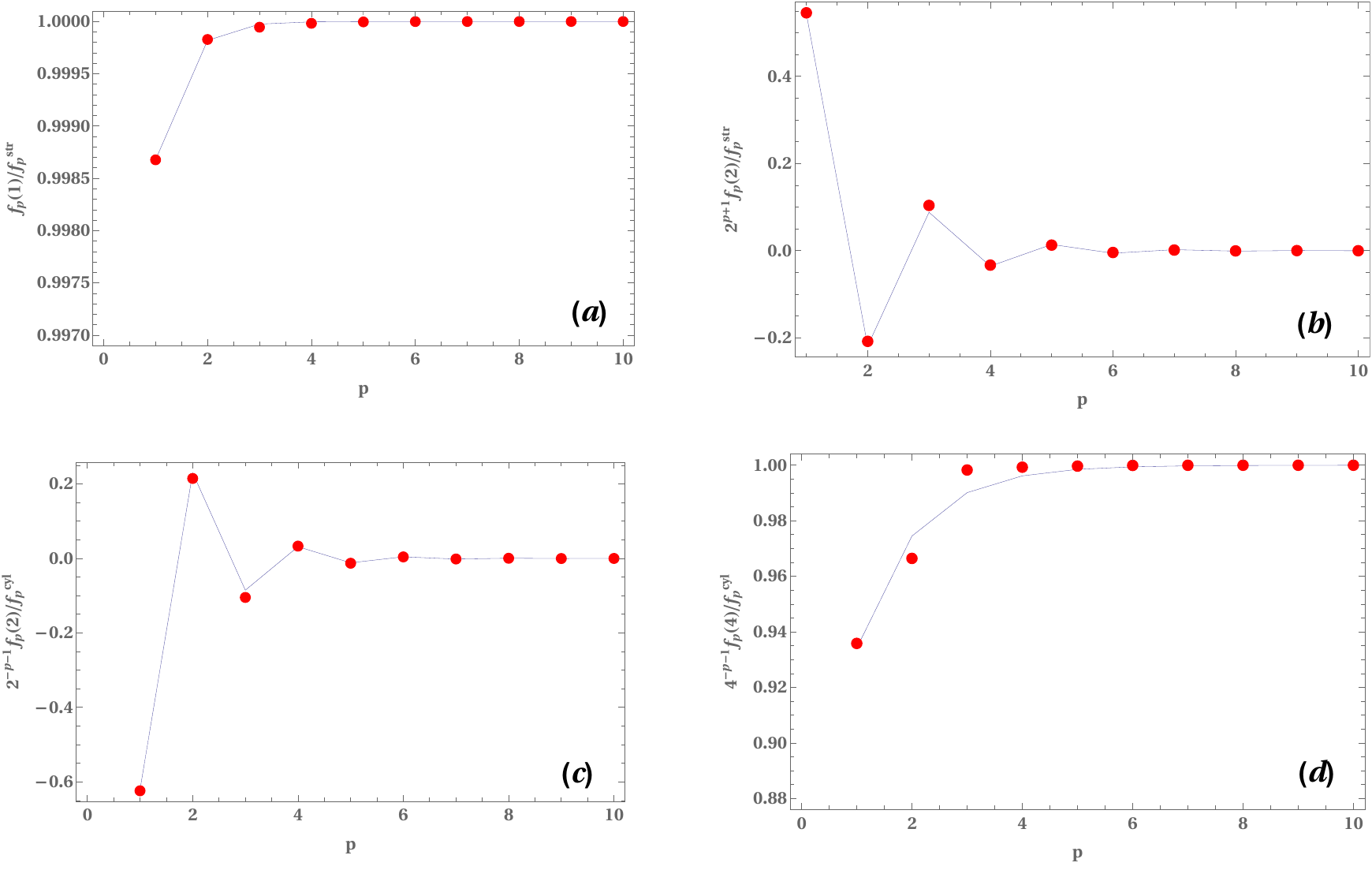}
	\caption{The behavior of ratio of the correction
		terms $f_p(\rho)$, as a function of $p$ for dimers on $(2M - 1) \times 2N$ cylinder. The dots represent our exact results. For $\rho = 1$ (a) the solid line is given by $a \, b^p + 1$, with $a=-0.00483$ and $b=0.76631$. For $\rho = 2$ the solid line is given by $(-1)^{p} a \, b^p$, with $a=-1.35106$ and $b=0.40254$ (b) and with $1.67625$ and $b=0.37047$ (c). For $\rho = 4$  (d) the solid line is given by $a \, b^p + 1$, with $a=-0.17005$ and $b=0.38714$. }
	\label{ratio_2}
\end{figure}

\begin{figure}[htt!]
  	\includegraphics[width=130mm]{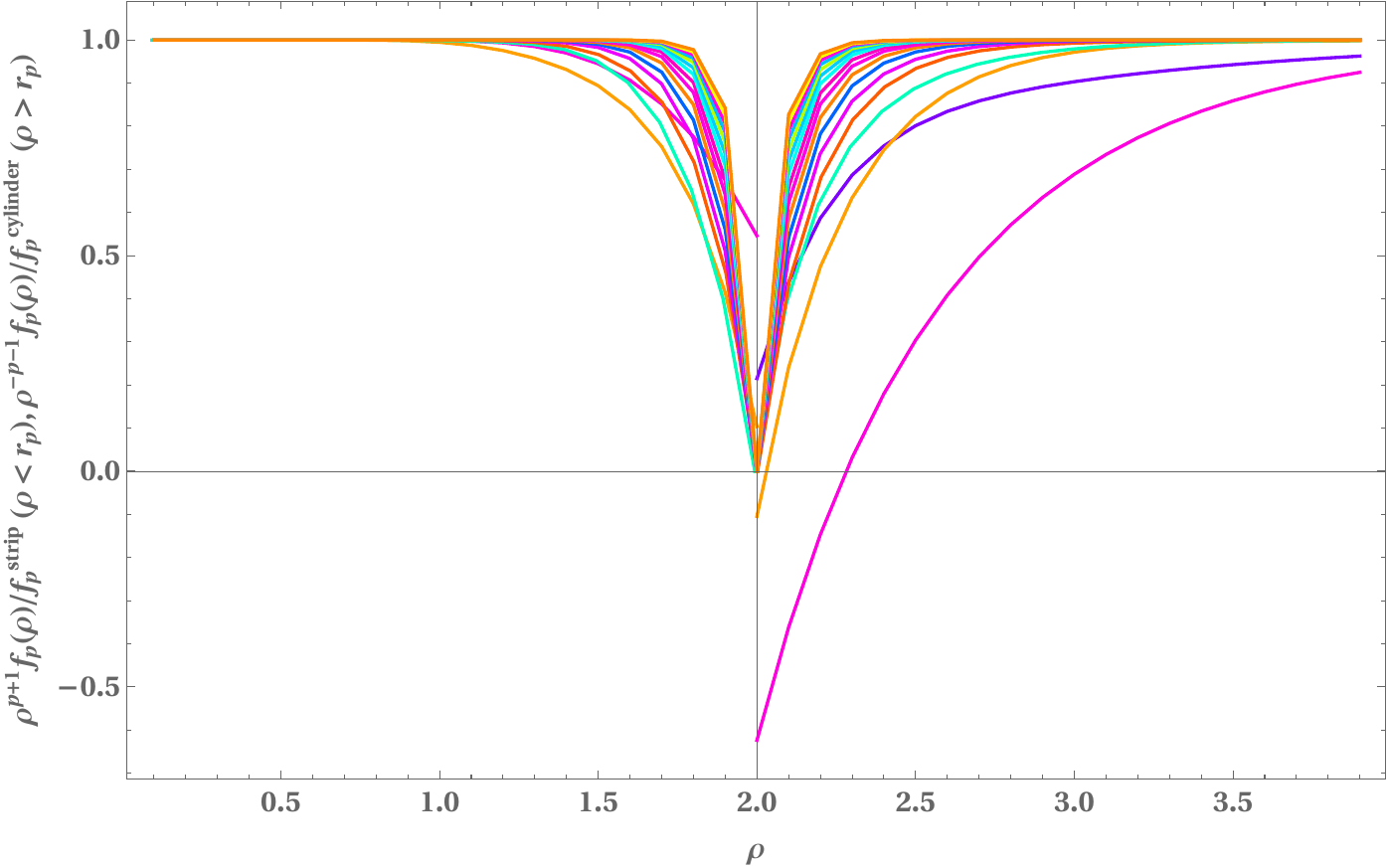}
	\caption{The behavior of ratios of the coefficients $f_p(\rho)$ for dimers on $(2M - 1) \times 2N$ cylinder in the asymptotic expansion of the free energy  with aspect ratio $\rho < 2$ times $\rho^{p+1}$  to the asymptotic coefficients for the strip $f_p^\mathrm{str}$ and the coefficients $f_p(\rho)$ with aspect ratio $\rho > 2$ times $\rho^{-p-1}$ to the asymptotic coefficients for the cylinder $f_p^\mathrm{cyl}$, as a function of $\rho$ for p =1, 2, 3,...,17.}
	\label{f_rho_2}
\end{figure}

For the dimers on $(2M - 1) \times 2N$ cylinder $\rho_{0} = 2$.
The values of $r_p^\mathrm{str}(\rho)$ and $r_p^\mathrm{cyl}(\rho)$ for $p=0$ up to $p=17$ are given in Table\ \ref{tabr2} and plotted them against $p$ in  Fig.\ \ref{ratio_2}. The dependence of the ratios  on $\rho$ is shown in Fig.\ \ref{f_rho_2}.
The behavior of the ratios is so different from what was in the previous case on the $2M \times 2N$ cylinder. Ratios $r_p^\mathrm{str}(\rho)$ and $r_p^\mathrm{cyl}(\rho)$ exponentially tend to $1$ in both cases as $p$ increase and $\rho\neq 2$. From  Table\ \ref{tabr2} and  Figs.\ \ref{ratio_2} and \ \ref{f_rho_2}, one sees also that the ratio $r_p$ exponentially tends to $0$ for $\rho = 2$ and  large values of $p$.
\begin{table}[ht!]
	\caption{Ratios of the coefficients in the asymptotic expansion of the free energy $f_p(\rho)$ for dimers on $(2M - 1) \times 2N$ cylinder with $\rho = 1$ to the asymptotic coefficients for the strip $f_p^\mathrm{str}$, $f_p(\rho)$ with $\rho = 2$ times $2^{p + 1}$ to the asymptotic coefficients for the strip $f_p^\mathrm{str}$ , $f_p(\rho)$ with $\rho = 2$ times $2^{-p-1}$ to the asymptotic coefficients for the infinite cylinder $f_p^\mathrm{cyl}$ and $f_p(\rho)$ with $\rho = 4$ times $4^{-p-1}$ to the asymptotic coefficients for the infinite cylinder $f_p^\mathrm{cyl}$ as a function of $p$.}
	\label{tabr2}
	\begin{center}
		\begin{tabular}{| c | c | c | c | c |}
			\hline
			$p$ & $r_p^\mathrm{str}(1)=\frac{f_p(1)}{f_{p}^\mathrm{str}}$ & $r_p^\mathrm{str}(2)=\frac{2^{p + 1} f_p(2)}{f_{p}^\mathrm{str}}$ & $r_p^\mathrm{cyl}(2)=\frac{2^{-p-1} f_p(2)}{f_{p}^\mathrm{cyl}}$ & $r_p^\mathrm{cyl}(4)=\frac{4^{-p-1} f_p(4)}{f_{p}^\mathrm{cyl}}$ \\\hline
			
            0 & -0.165476700114... & -0.992860200687... & 1.98572040137... & 1.99998667931...
\\\hline
			1 & 0.999163017403... & 0.545911084601... & -0.623898382401... & 0.935847573601... \\\hline
			2 & 0.994393914924... & -0.208056450107... & 0.214767948497... & 0.966455768238... \\\hline
			3 & 0.994617608672... & 0.103897509217... & -0.104715599840... &  0.998284327945... \\\hline
			4 & 0.998677187269... & -0.0331005405631... & 0.0331653165720... & 0.999298713663... \\\hline
			5 & 0.999828163357... & 0.0128382771942... & -0.0128445489466... & 0.999689579883... \\\hline
			6 & 0.999945502734... & -0.00413902974531... & 0.00413953505965... & 0.999931912363... \\\hline
			7 & 0.999982359481... & 0.00147927841899... & -0.00147932356436... & 0.999987167629... \\\hline
			8 & 0.999996073245... & -0.000479879907330... & 0.000479883568551... & 0.999996293657... \\\hline
			9 & 0.999999124863... & 0.000164673219211... & -0.000164673533301... & 0.999998968098... \\\hline
			10 & 0.999999761774... & -0.0000535304768250... & 0.0000535305023503... & 0.999999761363... \\\hline
			11 & 0.999999937979... & 0.0000179498954708... & -0.0000179498976106... & 0.999999942814... \\\hline
			12 & 0.999999985190... & $-0.583278564384...\times 10^{-7}$ & $0.58327858177...\times 10^{-7}$ & 0.999999985007... \\\hline
			13 & 0.999999996345... & $0.192834675475...\times 10^{-7}$ & $-0.19283467691192...\times 10^{-7}$ & 0.999999996204... \\\hline
			14 & 0.999999999063... & $-0.625668289807...\times 10^{-8}$ & $0.625668290972...\times 10^{-8}$ & 0.999999999074... \\\hline
			15 & 0.99999999977... & $0.204950099949...\times 10^{-8}$ & $-0.204950100045...\times 10^{-8}$ & 0.999999999769... \\\hline
			16 & 0.99999999994... & $-0.663683346672...\times 10^{-9}$ & $0.663683346749...\times 10^{-9}$ & 0.999999999942... \\\hline
			17 & 0.99999999999... & $0.216034307703...\times 10^{-9}$ & $-0.216034307709...\times 10^{-9}$ & 0.999999999985... \\\hline
		\end{tabular}
	\end{center}
\end{table}

\begin{figure}[htt!]
  	\includegraphics[width=130mm]{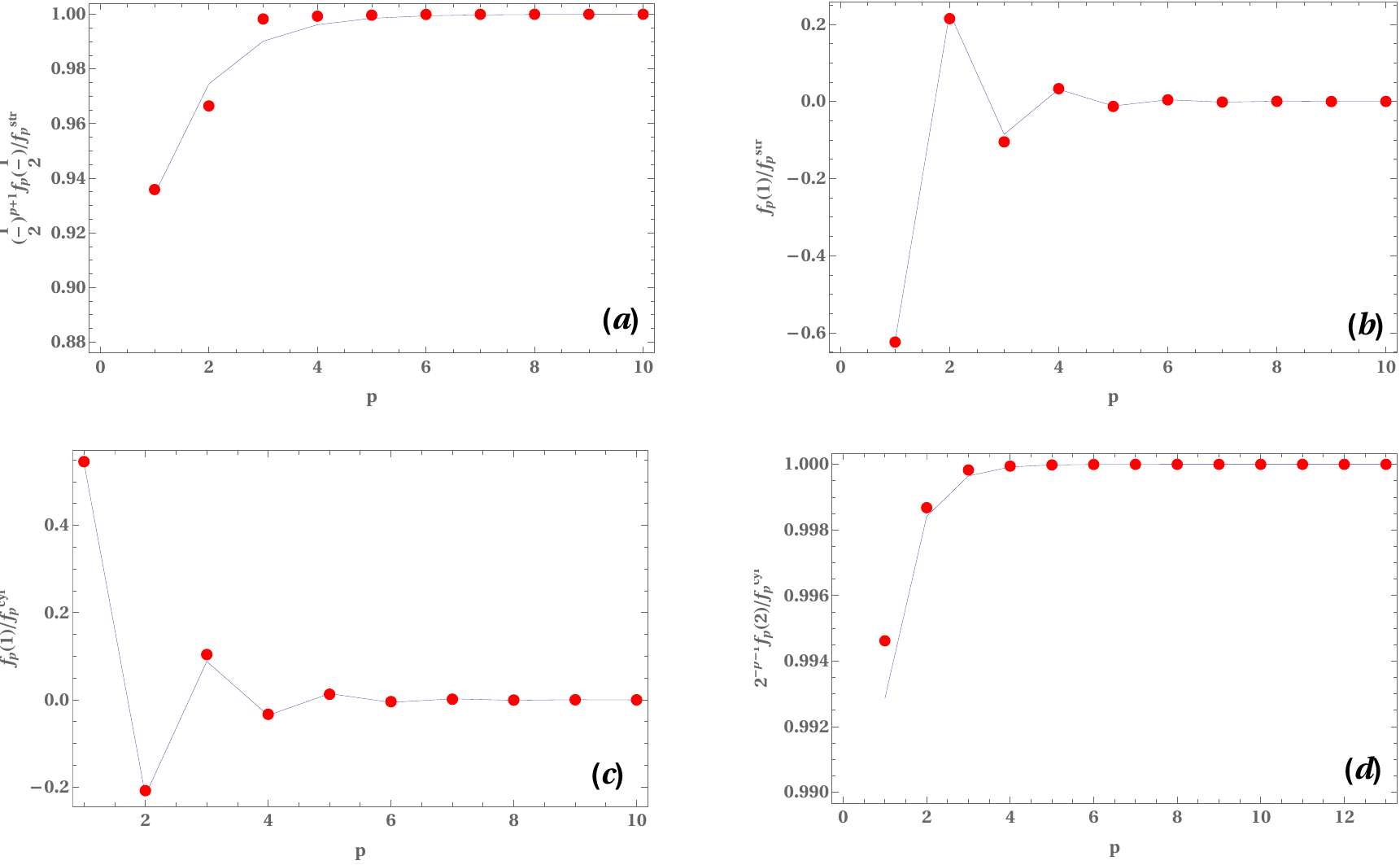}
	\caption{The behavior of ratio of the correction
		terms $f_p(\rho)$, as a function of $p$ for dimers on $2M \times (2N - 1)$ cylinder. The dots represent our exact results. For $\rho = 1/2$ (a) the solid line is given by $a \, b^p + 1$, with $a=-0.17005$ and $b=0.38714$. For $\rho = 1$ the solid line is given by $(-1)^{p} a \, b^p$, with $a=1.67625$ and $b=0.37047$ (b) and with $-1.35106$ and $b=0.40254$ (c). For $\rho = 2$  (d) the solid line is given by $a \, b^p + 1$, with $a=-0.00483$ and $b=0.76631$. }
	\label{ratio_3}
\end{figure}

\begin{figure}[htt!]
  	\includegraphics[width=130mm]{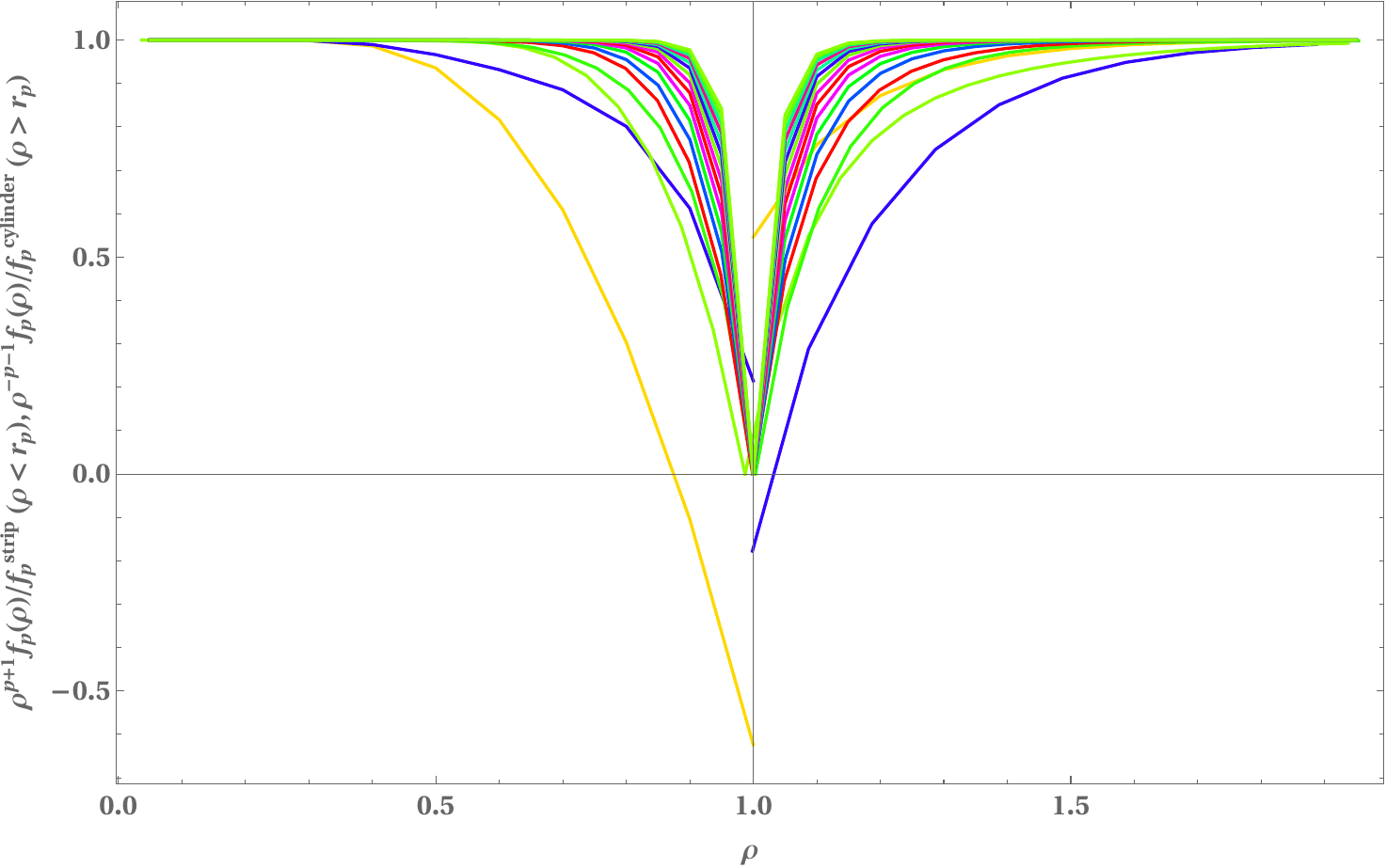}
	\caption{The behavior of ratios of the coefficients $f_p(\rho)$ for dimers on $2M \times (2N - 1)$ cylinder in the asymptotic expansion of the free energy  with aspect ratio $\rho < 1$ times $\rho^{p+1}$  to the asymptotic coefficients for the strip $f_p^\mathrm{str}$ and the coefficients $f_p(\rho)$ with aspect ratio $\rho > 1$ times $\rho^{-p-1}$ to the asymptotic coefficients for the cylinder $f_p^\mathrm{cyl}$, as a function of $\rho$ for p =1, 2, 3,...,17.}
	\label{f_rho_3}
\end{figure}

For the dimers on $2M \times (2N - 1)$ cylinder $\rho_{0} = 1$.
The results for the ratios $r_p^\mathrm{str}(\rho)$ and $r_p^\mathrm{cyl}(\rho)$ for $p=0$ up to $p=17$ are shown in Table\ \ref{tabr3} and Fig.\ \ref{ratio_3}. The ratios as a function of $\rho$ are demonstrated in Fig.\ \ref{f_rho_3}.
In this case, the behavior of the ratios is very similar to what was in the previous case on $(2M - 1) \times 2N$ cylinder, and differs only in the numerical values. Ratios $r_p^\mathrm{str}(\rho)$ and $r_p^\mathrm{cyl}(\rho)$ exponentially tend to $1$ in both cases as $p$ increase and $\rho\neq 1$. From  Table\ \ref{tabr3} and  Figs.\ \ref{ratio_3} and \ \ref{f_rho_3}, one also sees that the ratio  $r_p$ exponentially tends to $0$ for $\rho = 1$ and large values of $p$.

\begin{table}[ht!]
	\caption{Ratios of the coefficients in the asymptotic expansion of the free energy $f_p(\rho)$ for dimers on $2M \times (2N - 1)$ cylinder with $\rho = \frac{1}{2}$ times $(\frac{1}{2})^{p + 1}$ to the asymptotic coefficients for the strip $f_p^\mathrm{str}$, $f_p(\rho)$
 with $\rho = 1$ to the asymptotic coefficients for the strip $f_p^\mathrm{str}$, $f_p(\rho)$  with $\rho = 1$ to the asymptotic coefficients for the infinite cylinder $f_p^\mathrm{cyl}$, $f_p(\rho)$ and $f_p(\rho)$ with $\rho = 2$ times $2^{-p-1}$ to the asymptotic coefficients for the infinite cylinder $f_p^\mathrm{cyl}$ as a function of $p$.}
	\label{tabr3}
	\begin{center}
		\begin{tabular}{| c | c | c | c | c |}
			\hline
			$p$ & $r_p^\mathrm{str}(\frac{1}{2})=\frac{(\frac{1}{2})^{p + 1}f_p(\frac{1}{2})}{f_{p}^\mathrm{str}}$ & $r_p^\mathrm{str}(1)=\frac{f_p(1)}{f_{p}^\mathrm{str}}$ & $r_p^\mathrm{cyl}(1)=\frac{f_p(1)}{f_{p}^\mathrm{cyl}}$ & $r_p^\mathrm{cyl}(2)=\frac{2^{-p-1} f_p(2)}{f_{p}^\mathrm{cyl}}$ \\\hline
			
            0 & 0.992860200687... & 0.661906800458... & -0.330953400229... & 0.338086539197...
\\\hline
			1 & 0.935847573601... & -0.623898382401... & 0.545911084601... & 0.999163017403... \\\hline
			2 & 0.966455768238... & 0.214767948497... & -0.208056450107... & 0.994393914924... \\\hline
			3 & 0.998284327945... & -0.104715599840... & 0.103897509217... &  0.994617608672... \\\hline
			4 & 0.999298713663... & 0.0331653165720... & -0.0331005405631... & 0.998677187269... \\\hline
			5 & 0.999689579883... & -0.0128445489466... & 0.0128382771942... & 0.999828163357... \\\hline
			6 & 0.999931912363... & 0.00413953505965... & -0.00413902974531... & 0.999945502734... \\\hline
			7 & 0.999987167629... & -0.00147932356436... & 0.00147927841899... & 0.999982359481... \\\hline
			8 & 0.999996293657... & 0.000479883568551... & -0.000479879907330... & 0.999996073245... \\\hline
			9 & 0.999998968098... & -0.000164673533301... & 0.000164673219211... & 0.999999124863... \\\hline
			10 & 0.999999761363... & 0.0000535305023503... & -0.000053530476825... & 0.999999761774... \\\hline
			11 & 0.999999942814... & -0.0000179498976106... & 0.0000179498954708... & 0.999999937979... \\\hline
			12 & 0.999999985007... & $0.583278581767...\times 10^{-7}$ & $-0.583278564384...\times 10^{-7}$ & 0.999999985190... \\\hline
			13 & 0.999999996204... & $-0.192834676912...\times 10^{-7}$ & $0.192834675475...\times 10^{-7}$ & 0.999999996345... \\\hline
			14 & 0.999999999074... & $0625668290972...\times 10^{-8}$ & $-0.625668289807...\times 10^{-8}$ & 0.999999999063... \\\hline
			15 & 0.999999999769... & $-0204950100045...\times 10^{-8}$ & $0.204950099949...\times 10^{-8}$ & 0.99999999977... \\\hline
			16 & 0.999999999942... & $0.663683346749...\times 10^{-9}$ & $-0.663683346672...\times 10^{-9}$ & 0.99999999994... \\\hline
			17 & 0.999999999985... & $-0.216034307709...\times 10^{-9}$ & $0.216034307703...\times 10^{-9}$ & 0.99999999999... \\\hline
		\end{tabular}
	\end{center}
\end{table}

\section{Conclusions}
\label{concl}

The analysis of the finite-size corrections for the dimer model on a cylinder has been carried out for three different cases: $2M \times 2N$, $(2M  -1)\times 2N$ and $2M \times (2N - 1)$, when the lattice is completely covered by dimers. Based on results of \ \cite{Izmailian2003,Izmailian2002}
the  exact expressions for the correction terms $f_p(\rho)$ in the asymptotic expansion of the free energy have been derived up to $p=17$. Simple exact expressions for these correction terms have been obtained for two limiting cases: infinitely long strip ($N\rightarrow\infty$) and for the infinitely long cylinder ($M\rightarrow\infty$).

For $\rho<\rho_{0}$ the $r_p^\mathrm{str}(\rho)$ ratio of the coefficients $f_p(\rho)$ in the free energy expansion (multiplied by $\rho^{p+1}$) to the coefficients $f_p^\mathrm{str}$ for the infinitely long strip tends to $1$ as $p$ increases in all three cases. For $\rho >\rho_{0}$ $r_p^\text{cyl}$ the ratio of the coefficients $f_p(\rho)$ (multiplied by $\rho^{-p-1}$) to the coefficients $f_p^\mathrm{cyl}$ for the infinitely long cylinder also tends to $1$ as $p$ increases. For dimers on $2M \times 2N$ and $(2M  -1)\times 2N$ cylinders $\rho_{0} = 2$ and for dimer on $2M \times (2N - 1)$ cylinder $\rho_{0} = 1$.

Both the $r_p^\text{str}$ and $r_p^\text{cyl}$ ratios show an abrupt change at $\rho = \rho_{0}$ for all three types of cylinders.
For dimers on $2M \times 2N$ cylinder the ratios tends to $2$, and for dimers on $(2M  -1)\times 2N$ and $2M \times (2N - 1)$ cylinders, they tends to $0$.
Thus, we have similarities with the dimer case on a rectangle under free boundary conditions, as well as with the Ising model under Braskamp-Kunz boundary conditions in that there is an abrupt change. However, the limiting values of the ratios and the critical values of the aspect ratio at which this occurs differ.

As we can see in Fig.\ \ref{roots}, the critical values of the aspect ratio $\rho_{0}$ are equal to the limiting values of the roots of the correction terms $f_p(\rho)$, which for dimers on $(2M -1)\times 2N$ cylinder is equal to two, and for dimers on $2M \times (2N-1)$ cylinder is equal to one. For dimers on the $2M \times 2N$ cylinder, this critical value coincides with the point where $f_p(\rho)$ reaches its maximum at $\rho = 2$  (see Fig.\ \ref{fp_1}).

 The similarity in the behavior of the $r_p^{str}$ and $r_p^{cyl}$ ratios for the various models discussed in this paper and in references \ \cite{Izmailian2019, Izmailian2023}  appears to be due to the fact that the finite-size correction coefficients $f_p(\rho)$ are derived from Kronecker double series ${\rm K}_{2p+2}^{\alpha,\beta}$.

The $K_{2p+2}^{\frac{1}{2},\frac{1}{2}}(\rho)$ determines finite-size correction coefficients $f_p(\rho)$ for dimers on the $2M \times 2N$ cylinder and rectangle with free boundary conditions \cite{Izmailian2019}.

The $K_{2p+2}^{\frac{1}{2},0}(\rho)$ determines finite-size correction coefficients $f_p(\rho)$ for dimers on the $(2M -1)\times 2N$ cylinder and Ising model wtih Braskamp-Kunz boundary conditions \cite{Izmailian2023}.

The $K_{2p+2}^{0,\frac{1}{2}}(\rho)$ determines finite-size correction coefficients $f_p(\rho)$ for dimers on the $2M \times (2N-1)$ cylinder.

The dimer model is closely related to quantum spin systems, such as spin-1/2 chains, spin liquids, and valence bond solids \cite{Fradkin, Sachdev, Auerbach, Affleck, Rokhsar, Savary, Calvo}. Exact corrections to free energy could aid in predicting phase transitions or entanglement entropy in low-dimensional quantum materials, advancing the development of high-temperature superconductors, spintronic devices or quantum computing components. The exact results obtained serve as a benchmark for validating numerical methods, such as Monte Carlo simulations, in statistical mechanics. They also help to refine algorithms and ensure accuracy in predictive models for complex materials. The theoretical rigor of this study makes it possible to take into account the finite-size effects in the design of functional materials and the development of computational methods.

\section*{Acknowledgement}

It is with deep appreciation and gratitude that I remember my friend and colleague,
Nikolai Izmailian. He was one of the main driving forces behind this series of papers on finite correction factors. Sadly, he is no longer with us, and I had to complete this paper on my own.

{\bf Research data}

The source Wolfram Mathematica code that was used can be found on GitHub at https://github.com/vlpapoyan/cylinder\_code.
%%%%%%%%%%%%%%%%%%%%%%%%%%%%%%%%%%%%%%%%%%%%%%%%%%%%%%%%%%
\appendix
%%%%%%%%%%%%%%%%%%%%%%%%%%%%%
\section{Expressions of $f_p(\rho)$  for dimers on $2M \times 2N$ cylinder from $p=0$ up to $p=3$.}
\label{fexpression1}
%%%%%%%%%%%%%%%%%%%%%%%%%%%%%
\begin{eqnarray}
f_0 &=& -\frac{1}{3} \log \left(\frac{2 \theta _3^2}{\theta _2 \theta _4}\right)  \cr
f_1 &=& \frac{1}{48} \pi ^3 \rho ^2 \left(\frac{\theta _2^8}{30}-\frac{1}{2} \theta _3^2
   \theta _2^6+\frac{1}{30} \theta _4^4 \theta _2^4-\frac{1}{4} \theta _3^2
   \theta _4^4 \theta _2^2-\frac{7 \theta _4^8}{240}\right)  \cr
f_2 &=& \frac{\pi ^5 \rho ^3}{774144} \left(\theta _4^{12} \left(-42 \theta _2^2 \rho
   \text{E}+\theta _3^2 \rho  \left(62 \text{E}+105 \pi  \theta
   _2^2\right)-83 \pi  \theta _2^4 \rho -31\right)\right.\cr
   &-& \left.2 \theta _2^2 \theta
   _4^8 \left(105 \theta _3^4 \rho  \text{E}-3 \theta _3^2 \left(26
   \theta _2^2 \rho  \left(\text{E}+7 \pi  \theta
   _2^2\right)+21\right)+\theta _2^2 \left(1029 \theta _2^2 \rho
   \text{E}+34 \pi  \theta _2^4 \rho +39\right)\right)\right.\cr
   &-& \left.16 \theta _2^6
   \theta _4^4 \left(126 \theta _3^4 \rho  \text{E}-3 \theta _3^2
   \left(2 \theta _2^2 \rho  \text{E}+21 \pi  \theta _2^4 \rho
   +42\right)+\theta _2^2 \left(336 \theta _2^2 \rho  \text{E}+\pi
   \theta _2^4 \rho +3\right)\right)\right.\cr
   &-& \left.32 \theta _2^{10} \left(105 \theta
   _2^4 \rho  \text{E}+21 \theta _3^4 \rho  \text{E}-\theta _3^2
   \left(2 \theta _2^2 \rho  \text{E}+63\right)+\theta _2^2\right)-31
   \pi  \theta _4^{16} \rho \right)\cr
f_3 &=& \frac{\pi ^7 \rho ^4}{3715891200 \theta
   _3^2} \left(-70 \pi ^2 \rho ^2 \left(70 \theta _2^2+381
   \theta _3^2\right) \theta _4^{24}-70 \rho  \left(7035 \pi ^2 \rho
   \theta _2^6+1093 \pi ^2 \rho  \theta _3^2 \theta _2^4\right.\right.\cr
   &+& \left.\left.5 \left(6 \rho
   \text{E}^2-60 \pi  \rho  \theta _3^2 \text{E}+\pi  \left(41 \pi
   \rho  \theta _3^4+18\right)\right) \theta _2^2+254 \theta _3^2
   \left(\rho  \text{E} \left(\text{E}-6 \pi  \theta _3^2\right)+3
   \pi \right)\right) \theta _4^{20}\right.\cr
   &-& \left.\left(30099 \theta _3^2+70 \rho
   \left(5 \left(846 \rho  \text{E}^2+\pi  \left(5083 \pi  \rho  \theta
   _2^4+2502\right)\right) \theta _2^6+2 \left(407 \rho  \text{E}^2
   \theta _2^2\right.\right.\right.\right.\cr
   &+& \left.\left.\left.\left.\pi  \left(556 \pi  \rho  \theta _2^4+885\right) \theta
   _2^2-30 \text{E} \left(605 \pi  \rho  \theta _2^4+6\right)\right)
   \theta _3^2 \theta _2^2+10 \rho  \text{E} \left(127 \text{E}-96
   \pi  \theta _2^2\right) \theta _3^6\right.\right.\right.\cr
   &+& \left.\left.\left.\left(510 \rho  \text{E}^2
   \theta _2^2+5 \pi  \left(2423 \pi  \rho  \theta _2^4+108\right) \theta
   _2^2-12 \text{E} \left(295 \pi  \rho  \theta _2^4+127\right)\right)
   \theta _3^4\right)\right) \theta _4^{16}\right.\cr
   &-& \left.4 \theta _2^2 \left(175 \rho
   \left(2472 \rho  \text{E}^2+\pi  \left(3137 \pi  \rho  \theta
   _2^4+4914\right)\right) \theta _2^8+28 \left(5 \rho  \left(122 \rho
   \text{E}^2 \theta _2^2\right.\right.\right.\right.\cr
   &+& \left.\left.\left.\left.2 \pi  \left(29 \pi  \rho  \theta
   _2^4+81\right) \theta _2^2-45 \text{E} \left(380 \pi  \rho  \theta
   _2^4+277\right)\right) \theta _2^2+474\right) \theta _3^2 \theta
   _2^2+15750 \rho ^2 \text{E}^2 \theta _3^8\right.\right.\cr
   &+& \left.\left.140 \rho  \text{E}
   \left(266 \rho  \text{E} \theta _2^2-45 \left(248 \pi  \rho  \theta
   _2^4+3\right)\right) \theta _3^6+15 \left(7 \rho  \left(4165 \pi ^2 \rho
    \theta _2^6-432 \pi  \rho  \text{E} \theta _2^4\right.\right.\right.\right.\cr
   &+& \left.\left.\left.\left.90 \left(167 \rho
   \text{E}^2+93 \pi \right) \theta _2^2-448 \text{E}\right) \theta
   _2^2+474\right) \theta _3^4\right) \theta _4^{12}-16 \theta _2^6
   \left(525 \rho  \left(443 \rho  \text{E}^2\right.\right.\right.\cr
   &+& \left.\left.\left.17 \pi  \left(7 \pi  \rho
    \theta _2^4+30\right)\right) \theta _2^8+2 \left(711-70 \rho  \theta
   _2^2 \left(-2 \pi ^2 \rho  \theta _2^6+5190 \pi  \rho  \text{E}
   \theta _2^4+\left(47 \rho  \text{E}^2+3 \pi \right) \theta
   _2^2\right.\right.\right.\right.\cr
   &+& \left.\left.\left.\left.16875 \text{E}\right)\right) \theta _3^2 \theta _2^2+458850
   \rho ^2 \text{E}^2 \theta _3^8+140 \rho  \text{E} \left(23 \rho
   \text{E} \theta _2^2-45 \left(152 \pi  \rho  \theta
   _2^4+69\right)\right) \theta _3^6\right.\right.\cr
   &+& \left.\left.5 \left(7 \rho  \left(59145 \rho
   \text{E}^2 \theta _2^2+5 \pi  \left(293 \pi  \rho  \theta
   _2^4+3078\right) \theta _2^2+24 \text{E} \left(\pi  \rho  \theta
   _2^4-6\right)\right) \theta _2^2+49059\right) \theta _3^4\right) \theta
   _4^8\right.\cr
   &-& \left.32 \theta _2^{10} \left(2975 \rho  \left(36 \rho
   \text{E}^2+\pi  \left(\pi  \rho  \theta _2^4+18\right)\right) \theta
   _2^8-12 \left(35 \rho  \left(340 \pi  \rho  \text{E} \theta _2^4+2
   \left(9 \rho  \text{E}^2+\pi \right) \theta _2^2\right.\right.\right.\right.\cr
   &+& \left.\left.\left.\left.4335
   \text{E}\right) \theta _2^2+158\right) \theta _3^2 \theta
   _2^2+410550 \rho ^2 \text{E}^2 \theta _3^8-420 \rho  \text{E}
   \left(6 \rho  \text{E} \theta _2^2+85 \left(8 \pi  \rho  \theta
   _2^4+9\right)\right) \theta _3^6\right.\right.\cr
   &+& \left.\left.5 \left(7 \rho  \left(47430 \rho
   \text{E}^2 \theta _2^2-85 \pi  \left(\pi  \rho  \theta
   _2^4-54\right) \theta _2^2+48 \text{E} \left(\pi  \rho  \theta
   _2^4+4\right)\right) \theta _2^2+72522\right) \theta _3^4\right) \theta
   _4^4\right.\cr
   &+& \left.384 \theta _2^{14} \left(-2975 \rho ^2 \text{E}^2 \theta
   _2^8+\left(280 \rho  \text{E} \left(\rho  \text{E} \theta
   _2^2+255\right) \theta _2^2+79\right) \theta _3^2 \theta _2^2\right.\right.\cr
   &-& \left.\left.5 \left(7
   \rho  \text{E} \left(1955 \rho  \text{E} \theta _2^2+8\right)
   \theta _2^2+4029\right) \theta _3^4\right)\right)\cr
\cr
\nonumber
\end{eqnarray}

%%%%%%%%%%%%%%%%%%%%%%%%%%%%%%%%%%%%%%%%%%%%%%%%%%%
\section{Expressions of $f_p(\rho)$  for dimers on $(2M - 1) \times 2N$ cylinder from $p=0$ up to $p=3$.}
\label{fexpression2}
%%%%%%%%%%%%%%%%%%%%%%%%%%%%%
\begin{eqnarray}
f_0 &=& -\frac{1}{3} \log \left(\frac{2 \theta _4^2}{\theta _2 \theta _3}\right) \cr
f_1 &=& \frac{1}{48} \pi ^3 \rho ^2 \left(\frac{\theta _2^8}{30}+\frac{1}{2} \theta _3^2
   \theta _2^6+\frac{1}{30} \theta _4^4 \theta _2^4+\frac{1}{4} \theta _3^2
   \theta _4^4 \theta _2^2-\frac{7 \theta _4^8}{240}\right) \cr
f_2 &=& -\frac{\pi ^5 \rho ^3}{774144} \left(-3360 \theta _2^{14} \rho  \text{E}+16
   \theta _2^{12} \left(-4 \theta _3^2 \rho  \text{E}+\pi  \theta _4^4
   \rho +2\right)-336 \theta _2^{10} \left(\theta _4^4 \rho  \left(16
   \text{E}-\pi  \theta _3^2\right)\right.\right.\cr
   &-& \left.\left.2 \theta _3^2 \left(-\theta _3^2
   \rho  \text{E}+\pi  \theta _4^4 \rho +3\right)\right)+4 \theta _4^4
   \theta _2^8 \left(-24 \theta _3^2 \rho  \text{E}+17 \pi  \theta _4^4
   \rho +12\right)\right.\cr
   &+& \left.42 \theta _4^4 \theta _2^6 \left(-32 \theta _3^4 \rho
   \text{E}+2 \theta _3^2 \left(-8 \theta _3^2 \rho  \text{E}+9 \pi
    \theta _4^4 \rho +24\right)+\theta _4^4 \rho  \left(8 \pi  \theta
   _3^2-49 \text{E}\right)\right)\right.\cr
   &+& \left.\theta _4^8 \theta _2^4 \left(-156
   \theta _3^2 \rho  \text{E}+83 \pi  \theta _4^4 \rho +78\right)+21
   \theta _4^8 \theta _2^2 \left(-8 \theta _3^4 \rho  \text{E}+\theta
   _3^2 \left(-2 \theta _3^2 \rho  \text{E}+4 \pi  \theta _4^4 \rho
   +6\right)\right.\right.\cr
   &+& \left.\left.\theta _4^4 \rho  \left(\pi  \theta _3^2-2
   \text{E}\right)\right)+31 \theta _4^{12} \left(-2 \theta _3^2 \rho
   \text{E}+\pi  \theta _4^4 \rho +1\right)\right) \cr
f_3 &=& \frac{\pi ^7 \rho ^4}{11147673600 \theta _3^2} \left(571200 \pi ^2 \rho ^2 \theta _4^4 \theta
   _2^{22}+8960 \pi ^2 \rho ^2 \theta _3^2 \theta _4^4 \theta
   _2^{20}+285600 \pi  \rho  \theta _4^4 \left(\pi  \rho  \left(5 \theta
   _3^4+14 \theta _4^4\right)\right.\right.\cr
   &+& \left.\left.18\right) \theta _2^{18}+128 \theta _3^2
   \left(35 \pi  \rho  \left(-2 \pi  \rho  \theta _3^4+3 \pi  \rho  \theta
   _4^4+18\right) \theta _4^4+711\right) \theta _2^{16}+120 \left(65625 \pi
   ^2 \rho ^2 \theta _4^{12}\right.\right.\cr
   &+& \left.\left.70 \pi  \rho  \left(803 \pi  \rho  \theta
   _3^4+1530\right) \theta _4^8-2380 \pi  \rho  \theta _3^4 \left(7 \pi
   \rho  \theta _3^4-54\right) \theta _4^4+193392 \theta _3^4\right) \theta
   _2^{14}\right.\cr
   &-& \left.32 \theta _3^2 \theta _4^4 \left(35 \pi  \rho  \theta _4^4
   \left(16 \pi  \rho  \theta _3^4+77 \pi  \rho  \theta
   _4^4-18\right)-5688\right) \theta _2^{12}+60 \theta _4^4 \left(100940
   \pi ^2 \rho ^2 \theta _4^{12}\right.\right.\cr
   &+& \left.\left.735 \pi  \rho  \left(197 \pi  \rho  \theta
   _3^4+234\right) \theta _4^8+420 \pi  \rho  \theta _3^4 \left(1026-119
   \pi  \rho  \theta _3^4\right) \theta _4^4+580176 \theta _3^4\right)
   \theta _2^{10}\right.\cr
   &-& \left.16 \theta _3^2 \theta _4^8 \left(35 \pi  \rho  \left(-12
   \pi  \rho  \theta _3^4+457 \pi  \rho  \theta _4^4+486\right) \theta
   _4^4+4266\right) \theta _2^8+30 \theta _4^8 \left(54145 \pi ^2 \rho ^2
   \theta _4^{12}\right.\right.\cr
   &+& \left.\left.35 \pi  \rho  \left(3257 \pi  \rho  \theta
   _3^4+2502\right) \theta _4^8+210 \pi  \rho  \theta _3^4 \left(1674-161
   \pi  \rho  \theta _3^4\right) \theta _4^4+392472 \theta _3^4\right)
   \theta _2^6\right.\cr
   &-& \left.28 \theta _3^2 \theta _4^{12} \left(5 \pi  \rho  \left(-112
   \pi  \rho  \theta _3^4+1815 \pi  \rho  \theta _4^4+2655\right) \theta
   _4^4+5688\right) \theta _2^4+30 \theta _4^{12} \left(525 \pi ^2 \rho ^2
   \theta _4^{12}\right.\right.\cr
   &+& \left.\left.35 \pi  \rho  \left(47 \pi  \rho  \theta _3^4+18\right)
   \theta _4^8+35 \pi  \rho  \theta _3^4 \left(108-7 \pi  \rho  \theta
   _3^4\right) \theta _4^4+2844 \theta _3^4\right) \theta _2^2\right.\cr
   &-& \left.127 \theta
   _3^2 \theta _4^{16} \left(70 \pi  \rho  \left(-\pi  \rho  \theta _3^4+10
   \pi  \rho  \theta _4^4+18\right) \theta _4^4+711\right)+280 \rho ^2
   \text{E}^2 \left(8160 \theta _2^{22}+1152 \theta _3^2 \theta
   _2^{20}\right.\right.\cr
   &+& \left.\left.4080 \left(70 \theta _3^4+5 \theta _4^4\right) \theta
   _2^{18}+2560 \theta _3^2 \theta _4^4 \theta _2^{16}+60 \theta _4^4
   \left(9588 \theta _3^4+273 \theta _4^4\right) \theta _2^{14}+32 \left(35
   \theta _3^2 \theta _4^8\right.\right.\right.\cr
   &+& \left.\left.\left.28 \theta _3^6 \theta _4^4\right) \theta
   _2^{12}+30 \theta _4^4 \left(4896 \theta _3^8+11862 \theta _4^4 \theta
   _3^4+139 \theta _4^8\right) \theta _2^{10}-16 \theta _4^8 \left(36
   \theta _3^6+39 \theta _4^4 \theta _3^2\right) \theta _2^8\right.\right.\cr
   &+& \left.\left.30 \theta _4^8
   \left(2760 \theta _3^8+2219 \theta _4^4 \theta _3^4+\theta _4^8\right)
   \theta _2^6-\theta _4^{12} \left(1680 \theta _3^6+463 \theta _4^4 \theta
   _3^2\right) \theta _2^4+30 \theta _3^2 \theta _4^{12} \left(24 \theta
   _3^6\right.\right.\right.\cr
   &+& \left.\left.\left.12 \theta _4^4 \theta _3^2\right) \theta _2^2-127 \theta _3^2
   \theta _4^{16} \left(8 \theta _3^4+\theta _4^4\right)\right)-140 \rho
   \text{E} \left(128 \pi  \rho  \theta _4^4 \theta _2^{20}+4080
   \left(3 \left(5 \pi  \rho  \theta _4^4+42\right) \theta _3^2\right.\right.\right.\cr
   &+& \left.\left.\left.2 \left(8
   \pi  \rho  \theta _4^4+9\right) \theta _3^2\right) \theta _2^{18}+128
   \left(3 \pi  \rho  \theta _4^8+8 \pi  \rho  \theta _3^4 \theta _4^4+18
   \theta _3^4\right) \theta _2^{16}+120 \theta _4^4 \left(1394 \pi  \rho
   \theta _3^6\right.\right.\right.\cr
   &+& \left.\left.\left.51 \left(11 \pi  \rho  \theta _4^4+174\right) \theta _3^2+10
   \left(211 \pi  \rho  \theta _4^4+153\right) \theta _3^2\right) \theta
   _2^{14}+32 \theta _4^4 \left(5 \pi  \rho  \theta _4^8+\pi  \rho  \theta
   _3^4 \theta _4^4\right.\right.\right.\cr
   &+& \left.\left.\left.144 \theta _3^4\right) \theta _2^{12}+90 \theta _4^4
   \left(4 \left(793 \pi  \rho  \theta _4^4+612\right) \theta _3^6+\theta
   _4^4 \left(91 \pi  \rho  \theta _4^4+7362\right) \theta _3^2+18 \theta
   _4^4 \left(157 \pi  \rho  \theta _4^4\right.\right.\right.\right.\cr
   &+& \left.\left.\left.\left.91\right) \theta _3^2\right)
   \theta _2^{10}-16 \theta _4^8 \left(20 \pi  \rho  \theta _4^8+237 \pi
   \rho  \theta _3^4 \theta _4^4+108 \theta _3^4\right) \theta _2^8+15
   \theta _4^8 \left(6 \left(1327 \pi  \rho  \theta _4^4+1656\right) \theta
   _3^6\right.\right.\right.\cr
   &+& \left.\left.\left.\theta _4^4 \left(139 \pi  \rho  \theta _4^4+7470\right) \theta
   _3^2+18 \theta _4^4 \left(248 \pi  \rho  \theta _4^4+139\right) \theta
   _3^2\right) \theta _2^6-\theta _4^{12} \left(351 \pi  \rho  \theta
   _4^8+5086 \pi  \rho  \theta _3^4 \theta _4^4\right.\right.\right. \cr
   &+& \left. \left.\left.4032 \theta _3^4\right)
   \theta _2^4+15 \theta _4^{12} \left(\left(89 \pi  \rho  \theta
   _4^4+108\right) \theta _3^6+\theta _4^4 \left(\pi  \rho  \theta
   _4^4+18\right) \theta _3^2\right.\right.\right. \cr
   &+& \left.\left.\left.18 \left(2 \pi  \rho  \theta _4^8+\theta
   _4^4\right) \theta _3^2\right) \theta _2^2-127 \theta _4^{16} \left(18
   \theta _3^4+\pi  \rho  \theta _4^4 \left(17 \theta _3^4+\theta
   _4^4\right)\right)\right)\right) \cr
\cr
\nonumber
\end{eqnarray}

%%%%%%%%%%%%%%%%%%%%%%%%%%%%%%%%%%%%%%%%%%%%%%%%%%%
\section{Expressions of $f_p(\rho)$  for dimers on $2M \times (2N - 1)$ cylinder from $p=0$ up to $p=3$.}
\label{fexpression3}
%%%%%%%%%%%%%%%%%%%%%%%%%%%%%
\begin{eqnarray}
f_0 &=& -\frac{1}{6} \log \left(\frac{2 \theta _2^2}{\theta _3 \theta _4}\right) \cr
f_1 &=& \frac{1}{12} \pi ^3\rho ^2 \left(\frac{\theta _4^8}{30}+\frac{1}{30} \theta
   _4^4 \theta _2^4-\frac{7 \theta _2^8}{240}\right) \cr
f_2 &=& \frac{1}{144} \pi ^5 \rho ^3 \left(\frac{4}{3}
   \rho  \left(-\frac{31}{224} \theta _3^2 \theta
   _2^{12} \text{E}-\frac{13}{56} \theta _3^2
   \theta _4^4 \theta _2^8
   \text{E}+\frac{13}{56} \theta _4^3 \theta
   _2^8 \left(\frac{1}{4} \pi  \theta _3^4 \theta
   _4-\frac{1}{2} \theta _3^2 \theta _4
   \text{E}\right)\right.\right. \cr
   &-& \left.\left.\frac{1}{14} \theta _3^2
   \theta _4^8 \theta _2^4
   \text{E}+\frac{2}{7} \theta _4^7 \theta
   _2^4 \left(\frac{1}{4} \pi  \theta _3^4 \theta
   _4-\frac{1}{2} \theta _3^2 \theta _4
   \text{E}\right)+\frac{2}{7} \theta _4^{11}
   \left(\frac{1}{4} \pi  \theta _3^4 \theta
   _4-\frac{1}{2} \theta _3^2 \theta _4
   \text{E}\right)\right)\right. \cr
   &+& \left.4 \left(\frac{31
   \theta _2^{12}}{1344}+\frac{13}{224} \theta
   _4^4 \theta _2^8+\frac{1}{28} \theta _4^8
   \theta _2^4+\frac{\theta
   _4^{12}}{42}\right)\right) \cr
f_3 &=& \frac{\pi ^7 \rho ^4}{5760} \left(-\frac{40}{9} \left(-\frac{14}{5}
   \text{E} \theta _3^2 \theta _4^3 \left(\frac{1}{4} \pi  \theta
   _3^4 \theta _4-\frac{1}{2} \text{E} \theta _3^2 \theta
   _4\right) \theta _2^{12}+\frac{7}{120} \left(\frac{1}{4} \theta
   _3^4 \left(12 \text{E} \left(\text{E}-\pi  \theta
   _3^2\right)\right.\right.\right.\right. \cr
   &+& \left.\left.\left.\left.\pi ^2 \left(\theta _2^4+3 \theta _4^4\right)\right)
   \theta _4^4+12 \left(\frac{1}{4} \pi  \theta _3^4 \theta
   _4-\frac{1}{2} \text{E} \theta _3^2 \theta _4\right){}^2
   \theta _4^2\right) \theta _2^{12}-\frac{8}{5} \text{E} \theta
   _3^2 \theta _4^7 \left(\frac{1}{4} \pi  \theta _3^4 \theta
   _4\right.\right.\right. \cr
   &-& \left.\left.\left.\frac{1}{2} \text{E} \theta _3^2 \theta _4\right) \theta
   _2^8+\frac{1}{40} \left(\frac{1}{2} \theta _3^4 \left(12
   \text{E} \left(\text{E}-\pi  \theta _3^2\right)+\pi ^2
   \left(\theta _2^4+3 \theta _4^4\right)\right) \theta _4^8+56
   \left(\frac{1}{4} \pi  \theta _3^4 \theta _4\right.\right.\right.\right. \cr
   &-& \left.\left.\left.\left.\frac{1}{2}
   \text{E} \theta _3^2 \theta _4\right){}^2 \theta _4^6\right)
   \theta _2^8+\frac{16}{5} \text{E} \theta _3^2 \theta _4^{11}
   \left(\frac{1}{4} \pi  \theta _3^4 \theta _4-\frac{1}{2}
   \text{E} \theta _3^2 \theta _4\right) \theta _2^4-\frac{1}{15}
   \left(\frac{3}{4} \theta _3^4 \left(12 \text{E}
   \left(\text{E}-\pi  \theta _3^2\right)\right.\right.\right.\right.\cr
   &+& \left.\left.\left.\left.\pi ^2 \left(\theta
   _2^4+3 \theta _4^4\right)\right) \theta _4^{12}+132
   \left(\frac{1}{4} \pi  \theta _3^4 \theta _4-\frac{1}{2}
   \text{E} \theta _3^2 \theta _4\right){}^2 \theta
   _4^{10}\right) \theta _2^4+\frac{1}{15} \theta _4^{12}
   \left(\frac{1}{2} \theta _2^4 \theta _3^4 \left(\pi ^2 \theta
   _4^4-6 \text{E}^2\right)\right.\right.\right. \cr
   &-& \left.\left.\left.3 \text{E}^2 \theta _2^4 \theta
   _3^4\right)+\frac{1}{40} \theta _4^8 \left(14 \text{E}^2
   \theta _2^8 \theta _3^4-\theta _2^8 \theta _3^4 \left(\pi ^2
   \theta _4^4-6 \text{E}^2\right)\right)+\frac{7}{120} \theta
   _4^4 \left(33 \text{E}^2 \theta _2^{12} \theta
   _3^4\right.\right.\right. \cr
   &-& \left.\left.\left.\frac{3}{2} \theta _2^{12} \theta _3^4 \left(\pi ^2 \theta
   _4^4-6 \text{E}^2\right)\right)+\frac{127 \left(60
   \text{E}^2 \theta _2^{16} \theta _3^4-2 \theta _2^{16} \theta
   _3^4 \left(\pi ^2 \theta _4^4-6
   \text{E}^2\right)\right)}{3840}+\frac{1}{30} \left(-\theta
   _3^4 \left(12 \text{E} \left(\text{E}\right.\right.\right.\right.\right. \cr
   &-& \left.\left.\left.\left.\left.\pi  \theta
   _3^2\right)+\pi ^2 \left(\theta _2^4+3 \theta _4^4\right)\right)
   \theta _4^{16}-240 \left(\frac{1}{4} \pi  \theta _3^4 \theta
   _4-\frac{1}{2} \text{E} \theta _3^2 \theta _4\right){}^2
   \theta _4^{14}\right)\right) \rho ^2\right. \cr
   &-& \left.40 \left(-\frac{127}{480}
   \text{E} \theta _3^2 \theta _2^{16}-\frac{7}{20} \text{E}
   \theta _3^2 \theta _4^4 \theta _2^{12}+\frac{7}{30} \theta _4^3
   \left(\frac{1}{4} \pi  \theta _3^4 \theta _4-\frac{1}{2}
   \text{E} \theta _3^2 \theta _4\right) \theta
   _2^{12}-\frac{1}{10} \text{E} \theta _3^2 \theta _4^8 \theta
   _2^8\right.\right. \cr
   &+& \left.\left.\frac{1}{5} \theta _4^7 \left(\frac{1}{4} \pi  \theta _3^4
   \theta _4-\frac{1}{2} \text{E} \theta _3^2 \theta _4\right)
   \theta _2^8+\frac{2}{15} \text{E} \theta _3^2 \theta _4^{12}
   \theta _2^4-\frac{4}{5} \theta _4^{11} \left(\frac{1}{4} \pi
   \theta _3^4 \theta _4-\frac{1}{2} \text{E} \theta _3^2 \theta
   _4\right) \theta _2^4\right.\right. \cr
   &-& \left.\left.\frac{8}{15} \theta _4^{15}
   \left(\frac{1}{4} \pi  \theta _3^4 \theta _4-\frac{1}{2}
   \text{E} \theta _3^2 \theta _4\right)\right) \rho
   -\frac{632}{7} \left(\frac{127 \theta _2^{16}}{3840}+\frac{7}{120}
   \theta _4^4 \theta _2^{12}+\frac{1}{40} \theta _4^8 \theta
   _2^8\right.\right. \cr
   &-& \left.\left.\frac{1}{15} \theta _4^{12} \theta _2^4-\frac{\theta
   _4^{16}}{30}\right)\right) \cr
\cr
\nonumber
\end{eqnarray}

%%%%%%%%%%%%%%%%%%%%%%%%%%%%%%%%%%%
\section{Relation between elliptic theta functions, elliptic integral and gamma function}
\label{ThetaToGamma}
%%%%%%%%%%%%%%%%%%%%%%%%%%%%%
The elliptic theta functions and the elliptic integral of the second kind $E$ at particular values
of the aspect ratios $\rho = 1/2, 1, 2$ and $4$ given by
\begin{eqnarray}
\theta_2 = \frac{2^{7/8} \sqrt[8]{x}}{\sqrt[4]{\pi }}, \hskip 0.4cm
\theta_3 = \frac{\sqrt{2+\sqrt{2}} \sqrt[8]{x}}{\sqrt[4]{\pi }},\hskip 0.4cm
\theta_4 = \frac{\sqrt{2-\sqrt{2}}
   \sqrt[8]{x}}{\sqrt[4]{\pi }}, \nonumber\\
E = \frac{\sqrt{\pi } \sqrt[4]{x}}{2+\sqrt{2}}+\frac{\sqrt{\pi
   }}{\left(2+\sqrt{2}\right) \sqrt[4]{x}}
\end{eqnarray}
for $\rho = 1/2$,
\begin{eqnarray}
\theta_2 = \theta_4 = \sqrt[4]{\frac{2}{\pi }} \sqrt[8]{x}, \hskip 0.4cm
\theta_3 = \frac{\sqrt{2}}{\sqrt[4]{\pi }} \sqrt[8]{x}, \hskip 0.4cm
E = \frac{1}{4} \sqrt{\pi } \sqrt[4]{\frac{1}{x}}+\frac{\sqrt{\pi }}{2
   \sqrt[4]{\frac{1}{x}}}
\end{eqnarray}
for $\rho = 1$,
\begin{eqnarray}
\theta_2 = \frac{\sqrt{1-\frac{1}{\sqrt{2}}} \sqrt[8]{x}}{\sqrt[4]{\pi }},\hskip 0.4cm
\theta_3 = \frac{\sqrt{1+\frac{1}{\sqrt{2}}} \sqrt[8]{x}}{\sqrt[4]{\pi }},\hskip 0.4cm
\theta_4 =  \frac{2^{3/8} \sqrt[8]{x}}{\sqrt[4]{\pi }},\nonumber\\
E = \frac{\sqrt{\pi } \left(\sqrt{2}
   \sqrt{x}+\frac{1}{2+\sqrt{2}}\right)}{2 \sqrt[4]{x}}
\end{eqnarray}
for $\rho = 2$,
\begin{eqnarray}
\theta_2 = \frac{\left(\sqrt[4]{2}-1\right) \sqrt[8]{x}}{2^{3/4} \sqrt[4]{\pi }},\hskip 0.4cm
\theta_3 = \frac{\left(1+\sqrt[4]{2}\right) \sqrt[8]{x}}{2^{3/4} \sqrt[4]{\pi }},\hskip 0.4cm
\theta_4 = \frac{\sqrt[4]{\frac{2+\sqrt{2}}{\pi }} \sqrt[8]{x}}{\sqrt[16]{2}}, \nonumber\\
E = \frac{\sqrt{\pi } \left(\left(62+54 \sqrt[4]{2}+42 \sqrt{2}+34\
   2^{3/4}\right) \sqrt{x}+4\ 2^{3/4}+6 \sqrt{2}+4 \sqrt[4]{2}+3\right)}{2
   \left(1+\sqrt[4]{2}\right)^4 \left(2+\sqrt{2}+2\ 2^{3/4}\right)
   \sqrt[4]{x}}
\end{eqnarray}
for $\rho = 4$,
where
\begin{eqnarray}
\nonumber
x = \frac{\pi ^4}{16 \Gamma \left(\frac{3}{4}\right)^8} =   \frac{\Gamma[1/4]^8} {256 \pi^4} = 1.197316987373152280414040\ldots
\end{eqnarray}
and $\Gamma(z)$ is the gamma function.

%%%%%%%%%%%%%%%%%%%%%%%%%%%%%%%%%%%

%%%%%%%%%%%%%%%%%%%%%%%

%%%%%%%%%%%%%%%%%%%%%%%%%%%%%%%%%%%%%%%%%%%%%%%%%%%%%%%%%%%%%%%%%%%%%%%%%%%%%%%%%%%%

\end{document}

%%%%%%%%%%%%%%%%%%%%%%%%%%%%%%%%%%%%%%%%%%%%%%%%%%%%%%%%%

\begin{center}
\bf{SUPPLEMENTARY MATERIALS}
\end{center}
%%%%%%%%%%%%%%%%%%%%%%%%%%%%%%%%%%%%%%

\section{Kronecker functions ${\rm K}_{2p+2}^{0,\frac{1}{2}}(\tau)$ for $p$ from $1$ up to $17$}
\label{Kronecker}
\begin{eqnarray}
{\rm K}_{4}^{0,\frac{1}{2}}(\tau)&=& \frac{7 \theta _2^8}{240}-\frac{1}{30} \theta _4^4 \theta _2^4-\frac{\theta _4^8}{30} \cr
{\rm K}_{6}^{0,\frac{1}{2}}(\tau)&=& \frac{31 \theta _2^{12}}{1344}+\frac{13}{224} \theta _4^4 \theta
   _2^8+\frac{1}{28} \theta _4^8 \theta _2^4+\frac{\theta _4^{12}}{42} \cr
{\rm K}_{8}^{0,\frac{1}{2}}(\tau)&=& \frac{127 \theta _2^{16}}{3840}+\frac{7}{120} \theta _4^4 \theta
   _2^{12}+\frac{1}{40} \theta _4^8 \theta _2^8-\frac{1}{15} \theta _4^{12}
   \theta _2^4-\frac{\theta _4^{16}}{30} \frac{2555 \theta _2^{20}}{33792} \cr
   &+& \frac{3235 \theta _4^4 \theta
   _2^{16}}{16896}+\frac{335 \theta _4^8 \theta _2^{12}}{2112}+\frac{245
   \theta _4^{12} \theta _2^8}{1056}+\frac{25}{132} \theta _4^{16} \theta
   _2^4+\frac{5 \theta _4^{20}}{66} \cr
{\rm K}_{10}^{0,\frac{1}{2}}(\tau)&=& \frac{2555 \theta _2^{20}}{33792}+\frac{3235 \theta _4^4 \theta
   _2^{16}}{16896}+\frac{335 \theta _4^8 \theta _2^{12}}{2112}+\frac{245
   \theta _4^{12} \theta _2^8}{1056}+\frac{25}{132} \theta _4^{16} \theta
   _2^4+\frac{5 \theta _4^{20}}{66}  \cr
{\rm K}_{12}^{0,\frac{1}{2}}(\tau)&=& \frac{1414477 \theta _2^{24}}{5591040}+\frac{176639 \theta _4^4 \theta
   _2^{20}}{232960}+\frac{187423 \theta _4^8 \theta
   _2^{16}}{232960}-\frac{1753 \theta _4^{12} \theta
   _2^{12}}{10920} \cr
   &-& \frac{5191 \theta _4^{16} \theta
   _2^8}{7280}-\frac{691}{910} \theta _4^{20} \theta _2^4-\frac{691 \theta
   _4^{24}}{2730}  \cr
{\rm K}_{14}^{0,\frac{1}{2}}(\tau)&=& \frac{57337 \theta _2^{28}}{49152}+\frac{100345 \theta _4^4 \theta
   _2^{24}}{24576}+\frac{343}{64} \theta _4^8 \theta _2^{20}+\frac{217}{48}
   \theta _4^{12} \theta _2^{16}+\frac{665}{192} \theta _4^{16} \theta
   _2^{12}+\frac{175}{32} \theta _4^{20} \theta _2^8 \cr
   &+& \frac{49}{12} \theta
   _4^{24} \theta _2^4+\frac{7 \theta _4^{28}}{6}  \cr
{\rm K}_{16}^{0,\frac{1}{2}}(\tau)&=& \frac{118518239 \theta _2^{32}}{16711680}+\frac{3703687 \theta _4^4 \theta
   _2^{28}}{130560}+\frac{1157399 \theta _4^8 \theta
   _2^{24}}{26112}+\frac{493931 \theta _4^{12} \theta
   _2^{20}}{16320}+\frac{127963 \theta _4^{16} \theta
   _2^{16}}{32640} \cr
   &-& \frac{33979 \theta _4^{20} \theta
   _2^{12}}{1020}-\frac{9017}{204} \theta _4^{24} \theta
   _2^8-\frac{7234}{255} \theta _4^{28} \theta _2^4-\frac{3617 \theta
   _4^{32}}{510}  \cr
{\rm K}_{18}^{0,\frac{1}{2}}(\tau)&=& \frac{5749691557 \theta _2^{36}}{104595456}+\frac{4312268967 \theta _4^4
   \theta _2^{32}}{17432576}+\frac{488499261 \theta _4^8 \theta
   _2^{28}}{1089536}+\frac{32959793 \theta _4^{12} \theta
   _2^{24}}{77824} \cr
   &+& \frac{31281009 \theta _4^{16} \theta
   _2^{20}}{136192}+\frac{18818133 \theta _4^{20} \theta
   _2^{16}}{68096}+\frac{505069 \theta _4^{24} \theta
   _2^{12}}{1216}+\frac{1908813 \theta _4^{28} \theta
   _2^8}{4256}\cr
   &+& \frac{131601}{532} \theta _4^{32} \theta _2^4+\frac{43867
   \theta _4^{36}}{798} \cr
{\rm K}_{20}^{0,\frac{1}{2}}(\tau)&=& \frac{91546277357 \theta _2^{40}}{173015040}+\frac{11443284649 \theta _4^4
   \theta _2^{36}}{4325376}+\frac{715205289 \theta _4^8 \theta
   _2^{32}}{131072}+\frac{133613287 \theta _4^{12} \theta
   _2^{28}}{22528} \cr
   &+& \frac{160982011 \theta _4^{16} \theta
   _2^{24}}{45056}-\frac{672627 \theta _4^{20} \theta
   _2^{20}}{2560}-\frac{18741823 \theta _4^{24} \theta
   _2^{16}}{5632}-\frac{261889}{44} \theta _4^{28} \theta
   _2^{12} \cr
   &-& \frac{87303}{16} \theta _4^{32} \theta _2^8 - \frac{174611}{66}
   \theta _4^{36} \theta _2^4-\frac{174611 \theta _4^{40}}{330}  \cr
{\rm K}_{22}^{0,\frac{1}{2}}(\tau)&=& \frac{1792042792463 \theta _2^{44}}{289406976}+\frac{4928117679463 \theta
   _4^4 \theta _2^{40}}{144703488}+\frac{1424534016785 \theta _4^8 \theta
   _2^{36}}{18087936} \cr
   &+& \frac{298543652825 \theta _4^{12} \theta
   _2^{32}}{3014656}+\frac{27518308675 \theta _4^{16} \theta
   _2^{28}}{376832}+\frac{7765619543 \theta _4^{20} \theta
   _2^{24}}{188416} \cr
   &+& \frac{427130011 \theta _4^{24} \theta
   _2^{20}}{11776}+\frac{437010365 \theta _4^{28} \theta
   _2^{16}}{5888}+\frac{145695605 \theta _4^{32} \theta
   _2^{12}}{1472} \cr
   &+& \frac{173893775 \theta _4^{36} \theta
   _2^8}{2208}+\frac{9399643}{276} \theta _4^{40} \theta _2^4+\frac{854513
   \theta _4^{44}}{138} \cr
{\rm K}_{24}^{0,\frac{1}{2}}(\tau)&=& \frac{1982765468311237 \theta _2^{48}}{22900899840}+\frac{123922841769367
   \theta _4^4 \theta _2^{44}}{238551040}+\frac{317552282033747 \theta _4^8
   \theta _2^{40}}{238551040} \cr
   &+& \frac{4840399863055 \theta _4^{12} \theta
   _2^{36}}{2555904}+\frac{19393703261859 \theta _4^{16} \theta
   _2^{32}}{11927552}+\frac{184206451053 \theta _4^{20} \theta
   _2^{28}}{232960} \cr
   &+& \frac{9168090163 \theta _4^{24} \theta
   _2^{24}}{232960}-\frac{24235470579 \theta _4^{28} \theta
   _2^{20}}{29120}-\frac{37748942823 \theta _4^{32} \theta
   _2^{16}}{23296} \cr
   &-& \frac{590909935}{312} \theta _4^{36} \theta
   _2^{12}-\frac{4845463183 \theta _4^{40} \theta
   _2^8}{3640}-\frac{236364091}{455} \theta _4^{44} \theta
   _2^4-\frac{236364091 \theta _4^{48}}{2730}  \cr
{\rm K}_{26}^{0,\frac{1}{2}}(\tau)&=& \frac{286994504449393 \theta _2^{52}}{201326592}+\frac{932732139460537
   \theta _4^4 \theta _2^{48}}{100663296}+\frac{109304547593033 \theta _4^8
   \theta _2^{44}}{4194304} \cr
   &+& \frac{260511188624713 \theta _4^{12} \theta
   _2^{40}}{6291456}+\frac{499848052405 \theta _4^{16} \theta
   _2^{36}}{12288}+\frac{105766656879 \theta _4^{20} \theta
   _2^{32}}{4096} \cr
   &+& \frac{47882336177 \theta _4^{24} \theta
   _2^{28}}{4096}+\frac{26145301871 \theta _4^{28} \theta
   _2^{24}}{2048}+\frac{26109953727 \theta _4^{32} \theta
   _2^{20}}{1024} \cr
   &+& \frac{62519444845 \theta _4^{36} \theta
   _2^{16}}{1536}+\frac{7950109453}{192} \theta _4^{40} \theta
   _2^{12}+\frac{833927549}{32} \theta _4^{44} \theta
   _2^8 \cr
   &+& \frac{111190339}{12} \theta _4^{48} \theta _2^4+\frac{8553103
   \theta _4^{52}}{6}  \cr
{\rm K}_{28}^{0,\frac{1}{2}}(\tau)&=& \frac{3187598676787461083 \theta
   _2^{56}}{116769423360}+\frac{2789148842189028067 \theta _4^4 \theta
   _2^{52}}{14596177920} \cr
   &+& \frac{8541768329203897123 \theta _4^8 \theta
   _2^{48}}{14596177920}+\frac{117122372569551349 \theta _4^{12} \theta
   _2^{44}}{114032640} \cr
   &+& \frac{259109063470617689 \theta _4^{16} \theta
   _2^{40}}{228065280}+\frac{23286763093872943 \theta _4^{20} \theta
   _2^{36}}{28508160} \cr
   &+& \frac{3538174273793047 \theta _4^{24} \theta
   _2^{32}}{9502720}-\frac{1709050955003 \theta _4^{28} \theta
   _2^{28}}{148480}-\frac{213251182805423 \theta _4^{32} \theta
   _2^{24}}{593920} \cr
   &-& \frac{182462591800909 \theta _4^{36} \theta
   _2^{20}}{222720}-\frac{253011604133449 \theta _4^{40} \theta
   _2^{16}}{222720}-\frac{3574293880297 \theta _4^{44} \theta
   _2^{12}}{3480} \cr
   &-& \frac{4073032564951 \theta _4^{48} \theta
   _2^8}{6960}-\frac{166246227203}{870} \theta _4^{52} \theta
   _2^4-\frac{23749461029 \theta _4^{56}}{870}  \cr
{\rm K}_{30}^{0,\frac{1}{2}}(\tau)&=& \frac{4625594554880206790555 \theta
   _2^{60}}{7689065201664}+\frac{5781993193600258497145 \theta _4^4 \theta
   _2^{56}}{1281510866944} \cr
   &+& \frac{1197053278362553519085 \theta _4^8 \theta
   _2^{52}}{80094429184}+\frac{3449998116858039526685 \theta _4^{12} \theta
   _2^{48}}{120141643776} \cr
   &+& \frac{176832960797331398075 \theta _4^{16} \theta
   _2^{44}}{5005901824}+\frac{6593092025887844245 \theta _4^{20} \theta
   _2^{40}}{227540992} \cr
   &+& \frac{15057489582133109935 \theta _4^{24} \theta
   _2^{36}}{938606592}+\frac{1148059602135339555 \theta _4^{28} \theta
   _2^{32}}{156434432} \cr
   &+& \frac{270137147888555205 \theta _4^{32} \theta
   _2^{28}}{39108608}+\frac{950030979837219575 \theta _4^{36} \theta
   _2^{24}}{58662912} \cr
   &+& \frac{1608687356904545 \theta _4^{40} \theta
   _2^{20}}{55552}+\frac{10793174158056805 \theta _4^{44} \theta
   _2^{16}}{305536} \cr
   &+& \frac{13160697549795835 \theta _4^{48} \theta
   _2^{12}}{458304}+\frac{1141598969088565 \theta _4^{52} \theta
   _2^8}{76384} \cr
   &+& \frac{43079206380025 \theta _4^{56} \theta
   _2^4}{9548}+\frac{8615841276005 \theta _4^{60}}{14322}  \cr
{\rm K}_{32}^{0,\frac{1}{2}}(\tau)&=& \frac{16555640865486520478399 \theta
   _2^{64}}{1095216660480}+\frac{258681888523226882471 \theta _4^4 \theta
   _2^{60}}{2139095040} \cr
   &+& \frac{307184742621331922929 \theta _4^8 \theta
   _2^{56}}{713031680}+\frac{14146665707926859659 \theta _4^{12} \theta
   _2^{52}}{15728640} \cr
   &+& \frac{651393946463278680683 \theta _4^{16} \theta
   _2^{48}}{534773760}+\frac{6228887080973876947 \theta _4^{20} \theta
   _2^{44}}{5570560} \cr
   &+& \frac{11786578890080678351 \theta _4^{24} \theta
   _2^{40}}{16711680}+\frac{595381285712039003 \theta _4^{28} \theta
   _2^{36}}{2088960} \cr
   &+& \frac{16630389922136073 \theta _4^{32} \theta
   _2^{32}}{2785280}-\frac{19086340081578551 \theta _4^{36} \theta
   _2^{28}}{65280} \cr
   &-& \frac{45932939339157319 \theta _4^{40} \theta
   _2^{24}}{65280}-\frac{3041774757779101 \theta _4^{44} \theta
   _2^{20}}{2720} \cr
   &-& \frac{19878931911735149 \theta _4^{48} \theta
   _2^{16}}{16320}-\frac{26982623644207}{30} \theta _4^{52} \theta
   _2^{12}-\frac{73238549891519}{170} \theta _4^{56} \theta
   _2^8 \cr
   &-& \frac{30837284164868}{255} \theta _4^{60} \theta
   _2^4-\frac{7709321041217 \theta _4^{64}}{510} \cr
{\rm K}_{34}^{0,\frac{1}{2}}(\tau)&=& \frac{22142170099387402072897 \theta
   _2^{68}}{51539607552}+\frac{94104222922396458809825 \theta _4^4 \theta
   _2^{64}}{25769803776} \cr
   &+& \frac{11211635934113640600391 \theta _4^8 \theta
   _2^{60}}{805306368}+\frac{12636065094003310034501 \theta _4^{12} \theta
   _2^{56}}{402653184} \cr
   &+& \frac{4670767106947385967065 \theta _4^{16} \theta
   _2^{52}}{100663296}+\frac{2374534084792411879213 \theta _4^{20} \theta
   _2^{48}}{50331648} \cr
   &+& \frac{211501855244116702397 \theta _4^{24} \theta
   _2^{44}}{6291456}+\frac{53517891340676219947 \theta _4^{28} \theta
   _2^{40}}{3145728} \cr
   &+& \frac{5437171079376120883 \theta _4^{32} \theta
   _2^{36}}{786432}+\frac{2834234180556945355 \theta _4^{36} \theta
   _2^{32}}{393216} \cr
   &+& \frac{415226874730287961 \theta _4^{40} \theta
   _2^{28}}{24576}+\frac{413303620743459971 \theta _4^{44} \theta
   _2^{24}}{12288} \cr
   &+& \frac{144927536252353283 \theta _4^{48} \theta
   _2^{20}}{3072}+\frac{71270263627970839 \theta _4^{52} \theta
   _2^{16}}{1536} \cr
   &+& \frac{6025345368933245}{192} \theta _4^{56} \theta
   _2^{12}+\frac{1336531154563315}{96} \theta _4^{60} \theta
   _2^8 \cr
   &+& \frac{43820693592239}{12} \theta _4^{64} \theta
   _2^4+\frac{2577687858367 \theta _4^{68}}{6} \cr
{\rm K}_{36}^{0,\frac{1}{2}}(\tau)&=& \frac{904185845619475242495834469891 \theta
   _2^{72}}{65942866278481920}\cr
   &+& \frac{339069692107303215935937566361 \theta
   _4^4 \theta
   _2^{68}}{2747619428270080} \cr
   &+& \frac{21191855756706450995996072985 \theta
   _4^8 \theta _2^{64}}{42271068127232}\cr
   &+& \frac{26048322700402248267252707683
   \theta _4^{12} \theta
   _2^{60}}{21465776783360} \cr
   &+& \frac{11929472999561392648246013217 \theta
   _4^{16} \theta
   _2^{56}}{6133079080960}\cr
   &+& \frac{1663531642062914887257312417 \theta
   _4^{20} \theta _2^{52}}{766634885120} \cr
   &+& \frac{9245048420128302002459182253
   \theta _4^{24} \theta
   _2^{48}}{5366444195840}+\frac{11676973187927803183351401 \theta _4^{28}
   \theta _2^{44}}{11978670080} \cr
   &+& \frac{126205782930525638910647139 \theta
   _4^{32} \theta _2^{40}}{335402762240}-\frac{642965386584765688073407
   \theta _4^{36} \theta
   _2^{36}}{125776035840} \cr
   &-& \frac{15497212130327729663150577 \theta _4^{40}
   \theta _2^{32}}{41925345280}-\frac{91392309061448806209063 \theta
   _4^{44} \theta _2^{28}}{93583360} \cr
   &-& \frac{2256859065763943413441087 \theta
   _4^{48} \theta _2^{24}}{1310167040}-\frac{50767086757112251876401 \theta
   _4^{52} \theta _2^{20}}{23395840} \cr
   &-& \frac{45507327154426216630407 \theta
   _4^{56} \theta _2^{16}}{23395840}-\frac{776300510815076143111 \theta
   _4^{60} \theta _2^{12}}{639730} \cr
   &-& \frac{39472907329580193915 \theta
   _4^{64} \theta _2^8}{78736}-\frac{78945814659160432119 \theta _4^{68}
   \theta _2^4}{639730} \cr
   &-& \frac{26315271553053477373 \theta _4^{72}}{1919190}  \cr
\cr
\nonumber
\end{eqnarray}

%%%%%%%%%%%%%%%%%%%%%%%%%%%%%%%%%%%
\section{Expressions for $f_p(\rho)$  for $\rho = 1, 2, 4$ for dimers on $2M \times 2N$ cylinder}
\label{expressionfp1}
%%%%%%%%%%%%%%%%%%%%%%%%%%%%%%%%%%%
From now on
\begin{equation}
x = \frac{\pi^4}{16 \Gamma[3/4]^8} =  \frac{\Gamma[1/4]^8} {256 \pi^4} = 1.197316987373152280414040\ldots
\nonumber
\end{equation}
%%%%%%%%%%%%%%%%%%%%%%%%%%%%%%%%%%%

For $\rho=1$ one has:
\begin{eqnarray}
f_0 &=& -\frac{1}{3} \ln \left(2 \sqrt{2}\right) \cr
f_1 &=& \frac{1}{320} \left(1-20 \sqrt{2}\right) \pi  x \cr
f_2 &=& -\frac{\left(1+217 \sqrt{2}\right) \pi ^2 x^2}{5376} \cr
f_3 &=& -\frac{\pi ^3 x^2 \left(200 \left(1361 \sqrt{2}-1\right) x+42120
   \sqrt{2}+27\right)}{3686400} \cr
f_4 &=& -\frac{\pi ^4 x^3 \left(385 \left(40+478603 \sqrt{2}\right) x+324 \left(190971
   \sqrt{2}-17\right)\right)}{681246720} \cr
f_5 &=& \frac{\pi ^5 x^3}{27051687936000}\left(-2600 x \left(49 \left(347490379
   \sqrt{2}-3806\right) x\right.\right.\cr
   &+& \left.\left.162 \left(589+55464171
   \sqrt{2}\right)\right)-12252303 \left(91260
   \sqrt{2}-1\right)\right)\cr
f_6 &=& -\frac{\pi ^6 x^4}{784796221440} \left(77 \left(201712+150746382919 \sqrt{2}\right) x^2\right.\cr
   &+& \left.1188
   \left(7078289259 \sqrt{2}-9620\right) x+436671 \left(3+2241971
   \sqrt{2}\right)\right) \cr
 f_7 &=& \frac{\pi ^7 x^4}{58275828219248640000} \left(44200 x \left(-770 x \left(\left(319808588100347
   \sqrt{2}-54125992\right) x\right.\right.\right.\cr
   &+& \left.\left.\left.81 \left(612394+3646798000569
   \sqrt{2}\right)\right)-2616381 \left(17022995041
   \sqrt{2}-2881\right)\right)\right.\cr
   &-& \left.59089356711 \left(127+756281040
   \sqrt{2}\right)\right)\cr
f_8 &=& \frac{\pi ^8 x^5}{218842874912754892800}\cr
    &\times& \left(247 x \left(539 x \left(-35
   \left(3102279680+147501736849571663 \sqrt{2}\right) x\right.\right.\right.\cr
   &-& \left.\left.\left.648
   \left(8958278069833467 \sqrt{2}-189149944\right)\right)-12197628
   \left(1519120\right.\right.\right.\cr
   &+& \left.\left.\left.72228330717207 \sqrt{2}\right)\right)-36105786612
   \left(352980842919 \sqrt{2}-7453\right)\right)\cr
\cr
\nonumber
\end{eqnarray}
%%%%%%%%%%%%%%%%%%%%%%%%%%%%%%%%%%%

For $\rho = 2$ one has:
\begin{eqnarray}
f_0 &=& \frac{1}{24} \left(4 \log \left(10-7 \sqrt{2}\right)-5 \log (2)\right) \cr
f_1 &=& -\frac{\pi  x}{20} \cr
f_2 &=& -\frac{1}{84} \pi ^2 x^2 \cr
f_3 &=& -\frac{\pi ^3 x^2 (200 x+27)}{14400} \cr
f_4 &=& -\frac{\pi ^4 x^3 (3850 x+1377)}{166320} \cr
f_5 &=& -\frac{\pi ^5 x^3 (5200 x (93247 x+47709)+12252303)}{6604416000} \cr
f_6 &=& -\frac{\pi ^6 x^4 (176 x (88249 x+64935)+1310013)}{47900160} \cr
f_7 &=& -\frac{\pi ^7 x^4 (44200 x (1540 x (27062996
   x+24801957)+7537793661)+7504348302297)}{889218570240000} \cr
f_8 &=& -\frac{\pi ^8 x^5}{208704829132800}\cr
    &\times&\left(3952 x (539 x (1696559200
   x+1915143183)+289525947615)+67274106904809\right)\cr
\cr
\nonumber
\end{eqnarray}
%%%%%%%%%%%%%%%%%%%%%%%%%%%%%%%%%%%

For $\rho = 4$ one has:
\begin{eqnarray}
f_0 &=& \frac{1}{48} \left(4 \left(4 \log \left(\sqrt[4]{2}-1\right)-8 \log
   \left(1+\sqrt[4]{2}\right)+\log \left(2+\sqrt{2}\right)\right)-5 \log
   (2)\right) \cr
f_1 &=& \frac{1}{640} \left(2-40 \sqrt{2}\right) \pi  x \cr
f_2 &=& -\frac{\left(1+217 \sqrt{2}\right) \pi ^2 x^2}{5376} \cr
f_3 &=& \frac{\pi ^3 x^2}{7372800
   \left(1+\sqrt[4]{2}\right)^8 \left(3+4 \sqrt[4]{2}+6 \sqrt{2}+4\
   2^{3/4}\right)^{5/2}}\left(-400 \left(644261985\right.\right.\cr
   &+& \left.\left.539459470 \sqrt[4]{2}+451768984
   \sqrt{2}+381515938\ 2^{3/4}\right) x\right.\cr
   &-& \left.54 \left(739176193+618939026
   \sqrt[4]{2}+518333089 \sqrt{2}+437725248\ 2^{3/4}\right)\right)\cr
f_4 &=& \frac{\pi ^4 x^3}{170311680 \left(1+\sqrt[4]{2}\right)^{12}
   \left(2+\sqrt{2}+2\ 2^{3/4}\right)^3 \left(3+4 \sqrt[4]{2}+6 \sqrt{2}+4\
   2^{3/4}\right)^{7/2}}\cr
    &\times& \left(-385 \left(9270570753009931+7795938366639264
   \sqrt[4]{2}+6555701487994628 \sqrt{2}\right.\right.\cr
   &+& \left.\left.5512419345772896\ 2^{3/4}\right)
   x-162 \left(7397338084033110+6220673348640768
   \sqrt[4]{2}\right.\right.\cr
   &+& \left.\left.5231041575717583 \sqrt{2}+4398567341500128\
   2^{3/4}\right)\right) \cr
f_5 &=& \frac{\pi ^5 x^3}{6762921984000 \left(1+\sqrt[4]{2}\right)^{16}
   \left(2+\sqrt{2}+2\ 2^{3/4}\right)^4 \left(3+4 \sqrt[4]{2}+6 \sqrt{2}+4\
   2^{3/4}\right)^{11/2}}\cr
    &\times&  \left(2600 x \left(-49
   \left(552827529054787494507308+464870501568731627561308
   \sqrt[4]{2}\right.\right.\right.\cr
   &+& \left.\left.\left.390907932257939485768599 \sqrt{2}+328713210357764880552902\
   2^{3/4}\right) x\right.\right.\cr
   &-& \left.\left.162
   \left(88240107291500133831337+74200760236586255712714
   \sqrt[4]{2}\right.\right.\right.\cr
   &+& \left.\left.\left.62395152323461174463174 \sqrt{2}+52467880895348428417610\
   2^{3/4}\right)\right)\right.\cr
   &-& \left.12252303
   \left(145186868979113617505+122087068852942812246
   \sqrt[4]{2}\right.\right.\cr
   &+& \left.\left.102662576953552712119 \sqrt{2}+86328627457311172876\
   2^{3/4}\right)\right) \cr
f_6 &=&  {\scriptstyle \frac{\pi ^6 x^4}{196199055360 \left(1+\sqrt[4]{2}\right)^{20}
   \left(2+\sqrt{2}+2\ 2^{3/4}\right)^4 \left(3+4 \sqrt[4]{2}+6 \sqrt{2}+4\
   2^{3/4}\right)^8 \left(145+120 \sqrt[4]{2}+84 \sqrt{2}+72\
   2^{3/4}\right)^5}} \cr
    &\times& \scriptstyle \left(-77
   \left(569360602314200968951601009660316594431104644+47877328947270910804
   2725216557423518397266360
   \sqrt[4]{2}\right.\right.\cr
   &+& \left.\left.\scriptstyle 402598742836830049118036203994251760208299829
   \sqrt{2}+338543839637141240694702591175593961570905596\ 2^{3/4}\right)
   \scriptstyle{x^2}\right.\cr
   &-& \left.\scriptstyle 1188
   \left(26734249085542497340563305752477532102889768+224807342205326640528
   19750817222898405992008
   \sqrt[4]{2}\right.\right.\cr
   &+& \left.\left.\scriptstyle 18903968818317424556604963702094436016399177
   \sqrt{2}+15896279613391062526433869509570510148611964\ 2^{3/4}\right)
   \scriptstyle{x}\right.\cr
   &-& \left.\scriptstyle 436671
   \left(8467798259774775935456788799810801773581+7120541201736277304405473
   068428554169860 \sqrt[4]{2}\right.\right.\cr
   &+& \left.\left.\scriptstyle 5987637571206411823331003470687792163931
   \sqrt{2}+5034982969465911310243245638203296148184\
   2^{3/4}\right)\right) \cr
\cr
\nonumber
\end{eqnarray}
%%%%%%%%%%%%%%%%%%%%%%%%%%%%%%%%%%%
\section{The coefficients $f_{p}^{\,\mathrm {str}}$ for infinitely long strip
for dimers on $2M \times 2N$ cylinder.}
\label{fpstr1}
%%%%%%%%%%%%%%%%%%%%%%%%%%%%%%%%%%%

\begin{eqnarray}
f_{0}^{\mathrm {str}} &=& -\frac{\pi }{6} \cr
f_{1}^{\mathrm {str}} &=& -\frac{7 \pi ^3}{720} \cr
f_{2}^{\mathrm {str}} &=& -\frac{31 \pi ^5}{12096} \cr
f_{3}^{\mathrm {str}} &=& -\frac{10033 \pi ^7}{4838400} \cr
f_{4}^{\mathrm {str}} &=& -\frac{35989 \pi ^9}{10948608} \cr
f_{5}^{\mathrm {str}} &=& -\frac{180686706457 \pi ^{11}}{20922789888000} \cr
f_{6}^{\mathrm {str}} &=& -\frac{1248808051 \pi ^{13}}{36787322880} \cr
f_{7}^{\mathrm {str}} &=& -\frac{18201691489278203 \pi ^{15}}{97559980277760000} \cr
f_{8}^{\mathrm {str}} &=& -\frac{10510145076561543761 \pi ^{17}}{7693694821151539200} \cr
f_{9}^{\mathrm {str}} &=& -\frac{2641801043421964301500517 \pi ^{19}}{205531561650762547200000} \cr
f_{10}^{\mathrm {str}} &=& -\frac{890644111024794636680297 \pi ^{21}}{5893684261362126028800} \cr
\nonumber
\end{eqnarray}
%%%%%%%%%%%%%%%%%%%%%%%%%%%%%%%%%%%
\section{The coefficients $f_{p}^{\,\mathrm {cyl}}$ for infinitely long cylinder
for dimers on $2M \times 2N$ cylinder.}
\label{fpcyl1}
%%%%%%%%%%%%%%%%%%%%%%%%%%%%%%%%%%%

\begin{eqnarray}
f_{0}^{\mathrm {cyl}} &=& -\frac{\pi }{24} \cr
f_{1}^{\mathrm {cyl}} &=& -\frac{7 \pi ^3}{11520} \cr
f_{2}^{\mathrm {cyl}} &=& -\frac{31 \pi ^5}{774144} \cr
f_{3}^{\mathrm {cyl}} &=& -\frac{10033 \pi ^7}{1238630400} \cr
f_{4}^{\mathrm {cyl}} &=& -\frac{35989 \pi ^9}{11211374592} \cr
f_{5}^{\mathrm {cyl}} &=& -\frac{180686706457 \pi ^{11}}{85699747381248000} \cr
f_{6}^{\mathrm {cyl}} &=& -\frac{1248808051 \pi ^{13}}{602723498065920} \cr
f_{7}^{\mathrm {cyl}} &=& -\frac{18201691489278203 \pi ^{15}}{6393690867483279360000} \cr
f_{8}^{\mathrm {cyl}} &=& -\frac{10510145076561543761 \pi ^{17}}{2016855935195949092044800} \cr
f_{9}^{\mathrm {cyl}} &=& -\frac{2641801043421964301500517 \pi ^{19}}{215515462789509988692787200000} \cr
f_{10}^{\mathrm {cyl}} &=& -\frac{890644111024794636680297 \pi ^{21}}{24719903472168210651099955200} \cr
\nonumber
\end{eqnarray}
%%%%%%%%%%%%%%%%%%%%%%%%%%%%%%%%%%%
\section{Expressions for $f_p(\rho)$ for $\rho = 1, 2, 4$ for dimers on $(2M - 1) \times 2N$ cylinder}
\label{expressionfp2}
%%%%%%%%%%%%%%%%%%%%%%%%%%%%%%%%%%%

For $\rho = 1$ one has:
\begin{eqnarray}
f_0 &=& -\frac{\ln (2)}{4} \cr
f_1 &=& \frac{1}{320} \left(1+20 \sqrt{2}\right) \pi  x \cr
f_2 &=& \frac{\left(217 \sqrt{2}-1\right) \pi ^2 x^2}{5376} \cr
f_3 &=& \frac{\pi ^3 x^2 \left(200 \left(1+1361 \sqrt{2}\right) x+42120
   \sqrt{2}-27\right)}{3686400} \cr
f_4 &=& \frac{\pi ^4 x^3 \left(385 \left(478603 \sqrt{2}-40\right) x+324
   \left(17+190971 \sqrt{2}\right)\right)}{681246720} \cr
f_5 &=& \frac{\pi ^5 x^3}{27051687936000}\cr
    &\times& \left(2600 x \left(49 \left(3806+347490379 \sqrt{2}\right)
   x+162 \left(55464171 \sqrt{2}-589\right)\right)\right.\cr
   &+& \left.12252303 \left(1+91260
   \sqrt{2}\right)\right) \cr
f_6 &=& \frac{\pi ^6 x^4}{784796221440}\cr
    &\times&  \left(77 \left(150746382919 \sqrt{2}-201712\right)
   x^2+1188 \left(9620+7078289259 \sqrt{2}\right) x\right.\cr
   &+& \left.436671 \left(2241971
   \sqrt{2}-3\right)\right)\cr
f_7 &=& \frac{\pi ^7 x^4}{58275828219248640000} \cr
    &\times&   \left(44200 x \left(770 x
   \left(\left(54125992+319808588100347 \sqrt{2}\right) x \right.\right.\right.\cr
   &+& \left.\left.\left. 81
   \left(3646798000569 \sqrt{2}-612394\right)\right)+2616381
   \left(2881+17022995041 \sqrt{2}\right)\right)\right.\cr
   &+& \left. 59089356711
   \left(756281040 \sqrt{2}-127\right)\right) \cr
f_8 &=& \frac{\pi ^8 x^5}{218842874912754892800}  \cr
    &\times&    \left(247 x \left(539 x \left(35 \left(147501736849571663
   \sqrt{2}-3102279680\right) x\right.\right.\right.\cr
   &+& \left.\left.\left. 648 \left(189149944+8958278069833467
   \sqrt{2}\right)\right)+12197628 \left(72228330717207
   \sqrt{2}\right.\right.\right.\cr
   &-& \left.\left.\left. 1519120\right)\right)+36105786612 \left(7453+352980842919
   \sqrt{2}\right)\right) \cr
\cr
\nonumber
\end{eqnarray}
%%%%%%%%%%%%%%%%%%%%%%%%%%%%%%%%%%%

For $\rho = 2$ one has:
\begin{eqnarray}
f_0 &=& -\frac{3 \ln (2)}{4} \cr
f_1 &=& \frac{\pi  x}{80} \cr
f_2 &=& -\frac{1}{672} \pi ^2 x^2 \cr
f_3 &=& \frac{\pi ^3 x^2 (200 x-27)}{230400} \cr
f_4 &=& \frac{\pi ^4 (1377-3850 x) x^3}{5322240} \cr
f_5 &=& \frac{\pi ^5 x^3 (5200 x (93247 x-47709)+12252303)}{422682624000} \cr
f_6 &=& -\frac{\pi ^6 x^4 (176 x (88249 x-64935)+1310013)}{6131220480} \cr
f_7 &=& \frac{\pi ^7 x^4 (44200 x (1540 x (27062996
   x-24801957)+7537793661)-7504348302297)}{227639953981440000} \cr
f_8 &=& \frac{\pi ^8 x^5 (67274106904809-3952 x (539 x (1696559200
   x-1915143183)+289525947615))}{106856872515993600} \cr
\cr
\nonumber
\end{eqnarray}
%%%%%%%%%%%%%%%%%%%%%%%%%%%%%%%%%%%

For $\rho = 4$ one has:
\begin{eqnarray}
f_0 &=& \frac{1}{24} \left(-19 \log (2)-4 \log \left(2+\sqrt{2}\right)-8 \sinh
   ^{-1}(1)\right) \cr
f_1 &=& -\frac{3 \pi  x}{40} \cr
f_2 &=& -\frac{3}{56} \pi ^2 x^2 \cr
f_3 &=& -\frac{\pi ^3 x^2 (1000 x+153)}{9600} \cr
f_4 &=& -\frac{\pi ^4 x^3 (42350 x+14229)}{110880} \cr
f_5 &=& -\frac{\pi ^5 x^3 (5200 x (652729 x+344565)+85766121)}{1467648000} \cr
f_6 &=& -\frac{\pi ^6 x^4 (176 x (3794707 x+2748915)+56330559)}{31933440} \cr
\cr
\nonumber
\end{eqnarray}
%%%%%%%%%%%%%%%%%%%%%%%%%%%%%%%%%%%
\section{The coefficients $f_{p}^{\,\mathrm {str}}$ for infinitely long strip
for dimers on $(2M - 1) \times 2N$ cylinder.}
\label{fpstr2}
%%%%%%%%%%%%%%%%%%%%%%%%%%%%%%%%%%%
\begin{eqnarray}
f_{0}^{\mathrm {str}} &=& \frac{\pi }{3} \cr
f_{1}^{\mathrm {str}} &=& \frac{\pi ^3}{90} \cr
f_{2}^{\mathrm {str}} &=& \frac{\pi ^5}{378} \cr
f_{3}^{\mathrm {str}} &=& \frac{79 \pi ^7}{37800} \cr
f_{4}^{\mathrm {str}} &=& \frac{493 \pi ^9}{149688} \cr
f_{5}^{\mathrm {str}} &=& \frac{88269031 \pi ^{11}}{10216206000} \cr
f_{6}^{\mathrm {str}} &=& \frac{152461 \pi ^{13}}{4490640} \cr
f_{7}^{\mathrm {str}} &=& \frac{3888419459363 \pi ^{15}}{20841060240000} \cr
f_{8}^{\mathrm {str}} &=& \frac{80186655145391 \pi ^{17}}{58698233193600} \cr
f_{9}^{\mathrm {str}} &=& \frac{5038845219168059291 \pi ^{19}}{392020343114400000} \cr
f_{10}^{\mathrm {str}} &=& \frac{20809928059646128103 \pi ^{21}}{137706055072185600} \cr
\nonumber
\end{eqnarray}
%%%%%%%%%%%%%%%%%%%%%%%%%%%%%%%%%%%
\section{The coefficients $f_{p}^{\,\mathrm {cyl}}$ for infinitely long cylinder
for dimers on $(2M - 1) \times 2N$ cylinder.}
\label{fpcyl2}
%%%%%%%%%%%%%%%%%%%%%%%%%%%%%%%%%%%
\begin{eqnarray}
f_{0}^{\mathrm {cyl}} &=& -\frac{\pi }{24} \cr
f_{1}^{\mathrm {cyl}} &=& -\frac{7 \pi ^3}{11520} \cr
f_{2}^{\mathrm {cyl}} &=& -\frac{31 \pi ^5}{774144} \cr
f_{3}^{\mathrm {cyl}} &=& -\frac{10033 \pi ^7}{1238630400} \cr
f_{4}^{\mathrm {cyl}} &=& -\frac{35989 \pi ^9}{11211374592} \cr
f_{5}^{\mathrm {cyl}} &=& -\frac{180686706457 \pi ^{11}}{85699747381248000} \cr
f_{6}^{\mathrm {cyl}} &=& -\frac{1248808051 \pi ^{13}}{602723498065920} \cr
f_{7}^{\mathrm {cyl}} &=& -\frac{18201691489278203 \pi ^{15}}{6393690867483279360000} \cr
f_{8}^{\mathrm {cyl}} &=& -\frac{10510145076561543761 \pi ^{17}}{2016855935195949092044800} \cr
f_{9}^{\mathrm {cyl}} &=& -\frac{2641801043421964301500517 \pi ^{19}}{215515462789509988692787200000} \cr
f_{10}^{\mathrm {cyl}} &=& -\frac{890644111024794636680297 \pi ^{21}}{24719903472168210651099955200} \cr
\nonumber
\end{eqnarray}
%%%%%%%%%%%%%%%%%%%%%%%%%%%%%%%%%%%
\section{Expressions for $f_p(\rho)$  for $\rho = 1/2, 1, 2$
for dimers on $2M \times (2N - 1)$ cylinder}
\label{expressionfp3}
%%%%%%%%%%%%%%%%%%%%%%%%%%%%%%%%%%%

For $\rho = 1/2$ one has:
\begin{eqnarray}
f_0 &=& -\frac{3 \ln (2)}{8} \cr
f_1 &=& -\frac{3 \pi  x}{40} \cr
f_2 &=& -\frac{3}{28} \pi ^2 x^2 \cr
f_3 &=& -\frac{\pi ^3 x^2 (1000 x+153)}{2400} \cr
f_4 &=& -\frac{\pi ^4 x^3 (42350 x+14229)}{13860}\cr
f_5 &=& -\frac{\pi ^5 x^3 (5200 x (652729 x+344565)+85766121)}{91728000} \cr
f_6 &=& -\frac{\pi ^6 x^4 (176 x (3794707 x+2748915)+56330559)}{997920} \cr
f_7 &=& -\pi ^7 \left(\frac{115017733 x^7}{6804}+\frac{78692629
   x^6}{5040}-\frac{527335359 x^5}{172480}+\frac{4908285459
   x^4}{70720000}\right)\cr
f_8 &=& -\pi ^8 \left(\frac{46049464 x^8}{81}+\frac{1725993239
   x^7}{2700}+\frac{13582698777 x^6}{75460}+\frac{45827475939
   x^5}{4347200}\right) \cr
\cr
\nonumber
\end{eqnarray}
%%%%%%%%%%%%%%%%%%%%%%%%%%%%%%%%%%%

For $\rho = 1$ one has:
\begin{eqnarray}
f_0 &=& -\frac{\ln (2)}{8} \cr
f_1 &=& \frac{\pi  x}{80} \cr
f_2 &=& -\frac{1}{336} \pi ^2 x^2 \cr
f_3 &=& \frac{\pi ^3 x^2 (200 x-27)}{57600} \cr
f_4 &=& \frac{\pi ^4 (1377-3850 x) x^3}{665280} \cr
f_5 &=& \frac{\pi ^5 x^3 (5200 x (93247 x-47709)+12252303)}{26417664000} \cr
f_6 &=& -\frac{\pi ^6 x^4 (176 x (88249 x-64935)+1310013)}{191600640} \cr
f_7 &=& \frac{\pi ^7 x^4 (44200 x (1540 x (27062996
   x-24801957)+7537793661)-7504348302297)}{3556874280960000} \cr
f_8 &=& \pi ^8 \left(-\frac{302957 x^8}{69984}+\frac{23643743
   x^7}{4838400}-\frac{79430987 x^6}{57953280}+\frac{89681949
   x^5}{1112883200}\right) \cr
\cr
\nonumber
\end{eqnarray}
%%%%%%%%%%%%%%%%%%%%%%%%%%%%%%%%%%%

For $\rho = 2$ one has:
\begin{eqnarray}
f_0 &=& \frac{1}{24} \log \left(35+\frac{99}{2 \sqrt{2}}\right) \cr
f_1 &=& -\frac{3 \pi  x}{40} \cr
f_2 &=& -\frac{3}{28} \pi ^2 x^2 \cr
f_3 &=& -\frac{\pi ^3 x^2 (1000 x+153)}{2400} \cr
f_4 &=& -\frac{\pi ^4 x^3 (42350 x+14229)}{13860} \cr
f_5 &=& -\frac{\pi ^5 x^3 (5200 x (652729 x+344565)+85766121)}{91728000} \cr
f_6 &=& -\frac{\pi ^6 x^4 (176 x (3794707 x+2748915)+56330559)}{997920} \cr
f_7 &=& \pi ^7 \left(-\frac{115017733 x^7}{6804}-\frac{78692629
   x^6}{5040}-\frac{527335359 x^5}{172480}-\frac{4908285459
   x^4}{70720000}\right) \cr
f_8 &=& \pi ^8 \left(-\frac{46049464 x^8}{81}-\frac{1725993239
   x^7}{2700}-\frac{13582698777 x^6}{75460}-\frac{45827475939
   x^5}{4347200}\right) \cr
\cr
\nonumber
\end{eqnarray}
%%%%%%%%%%%%%%%%%%%%%%%%%%%%%%%%%%%
\section{The coefficients $f_{p}^{\,\mathrm {str}}$ for infinitely long strip
for dimers on $2M \times (2N - 1)$ cylinder.}
\label{fpstr3}
%%%%%%%%%%%%%%%%%%%%%%%%%%%%%%%%%%%
\begin{eqnarray}
f_{0}^{\mathrm {str}} &=& -\frac{\pi }{24} \cr
f_{1}^{\mathrm {str}} &=& -\frac{7 \pi ^3}{2880} \cr
f_{2}^{\mathrm {str}} &=& -\frac{31 \pi ^5}{48384} \cr
f_{3}^{\mathrm {str}} &=& -\frac{10033 \pi ^7}{19353600} \cr
f_{4}^{\mathrm {str}} &=& -\frac{35989 \pi ^9}{43794432} \cr
f_{5}^{\mathrm {str}} &=& -\frac{180686706457 \pi ^{11}}{83691159552000} \cr
f_{6}^{\mathrm {str}} &=& -\frac{1248808051 \pi ^{13}}{147149291520} \cr
f_{7}^{\mathrm {str}} &=& -\frac{18201691489278203 \pi ^{15}}{390239921111040000} \cr
f_{8}^{\mathrm {str}} &=& -\frac{10510145076561543761 \pi ^{17}}{30774779284606156800} \cr
f_{9}^{\mathrm {str}} &=& -\frac{2641801043421964301500517 \pi ^{19}}{822126246603050188800000} \cr
f_{10}^{\mathrm {str}} &=& -\frac{890644111024794636680297 \pi ^{21}}{23574737045448504115200} \cr
\nonumber
\end{eqnarray}
%%%%%%%%%%%%%%%%%%%%%%%%%%%%%%%%%%%
\section{The coefficients $f_{p}^{\,\mathrm {cyl}}$ for infinitely long cylinder
for dimers on $2M \times (2N - 1)$ cylinder.}
\label{fpcyl3}
%%%%%%%%%%%%%%%%%%%%%%%%%%%%%%%%%%%
\begin{eqnarray}
f_{0}^{\mathrm {cyl}} &=& \frac{\pi }{12} \cr
f_{1}^{\mathrm {cyl}} &=& \frac{\pi ^3}{360} \cr
f_{2}^{\mathrm {cyl}} &=& \frac{\pi ^5}{1512} \cr
f_{3}^{\mathrm {cyl}} &=& \frac{79 \pi ^7}{151200} \cr
f_{4}^{\mathrm {cyl}} &=& \frac{493 \pi ^9}{598752} \cr
f_{5}^{\mathrm {cyl}} &=& \frac{88269031 \pi ^{11}}{40864824000} \cr
f_{6}^{\mathrm {cyl}} &=& \frac{152461 \pi ^{13}}{17962560}\cr
f_{7}^{\mathrm {cyl}} &=& \frac{3888419459363 \pi ^{15}}{83364240960000} \cr
f_{8}^{\mathrm {cyl}} &=& \frac{80186655145391 \pi ^{17}}{234792932774400} \cr
f_{9}^{\mathrm {cyl}} &=& \frac{5038845219168059291 \pi ^{19}}{1568081372457600000}\cr
f_{10}^{\mathrm {cyl}} &=& \frac{20809928059646128103 \pi ^{21}}{550824220288742400} \cr
\nonumber
\end{eqnarray}

%%%%%%%%%%%%%%%%%%%%%%%%%%%%%%%%%%%%%%%%%%%%%%%%%%%%%%